\documentclass[iop]{emulateapj}
\usepackage{apjfonts}
\usepackage[bookmarks=false, colorlinks=true, citecolor=blue,
backref=true]{hyperref}

\newcommand{\sfr}{{\rm SFR}}
\newcommand{\msperyr}{{M_{\odot}{\rm yr}^{-1}}}
\newcommand{\ms}{{M_{\odot}}}
\newcommand{\ha}{H$\alpha$}
\newcommand{\mic}{\mu{\rm m}}

\shorttitle{STAR FORMATION RATE DISTRIBUTIONS}
\shortauthors{SALIM AND LEE}

\begin{document}


\title{Star Formation Rate Distributions: Inadequacy of the Schechter Function}

\author{Samir Salim} 
\affil{Department of Astronomy, Indiana University,
  Bloomington, IN 47404, salims@indiana.edu} 
\and
\author{Janice C.\ Lee\altaffilmark{1}} 
\affil{Space Telescope Science Institute, Baltimore, MD 21218} 
\altaffiltext{1}{Visiting Astronomer, Spitzer Science Center, Caltech, Pasadena, CA 91125.}


\begin{abstract}
  In this paper we posit that galaxy luminosity functions (LFs) come
  in two fundamentally different types depending on whether the
  luminosity traces galaxy stellar mass or its current star formation
  rate (SFR).  {\it Mass function types} reflect the older stars and
  therefore the stellar mass distribution, while {\it SFR function
    types} arise from the young stars and hence the distribution of
  SFRs. Optical and near-infrared LFs are of the mass function type,
  and are well fit by a Schechter function (power law with an
  exponential cutoff at the bright end). In contrast, LFs of the SFR
  function type are of a different form, one that cannot be adequately
  described by a Schechter function. We demonstrate this difference by
  generating SFR distributions for mock samples of galaxies drawn from
  a Schechter stellar mass distribution along with established
  empirical relations between the SFR and stellar mass. Compared with
  the Schechter function, SFR distributions have a shallower decline
  at the bright end, which can be traced to the large intrinsic
  scatter of SFRs at any given stellar mass. A superior description of
  SFR distributions is given by the ``Saunders'' function, which
  combines a power law with a Gaussian at the high end. We show that
  the Schechter-like appearance of UV and \ha\ LFs, although they are
  LFs of SFR function type, results when luminosities are not
  corrected for dust, or when average statistical corrections are used
  because individual attenuation measurements are not available. We
  thus infer that the non-Schechter form of the far-IR LFs is a true
  reflection of the underlying SFR distribution, rather than the
  purported artifact of AGN contamination.
\end{abstract}


\section{Introduction}

\citet{schechter} realized that the distribution of optical
luminosities of cluster galaxies empirically follows the same
functional form that has been introduced on theoretical grounds by
\citet{press_schechter} to describe the halo mass function. This
functional form is now known as the {\it Schechter function}. It
combines a power law at the faint end with an exponential cutoff at
the bright end and is uniquely determined with three parameters. The
function has been shown to describe luminosities of galaxies in field
environments too \citep{felten}. Parameterization of the optical
luminosity function (LF) using a Schechter function has simplified
comparisons of different samples of galaxies, including samples at
different redshifts, and the determination of the cosmic luminosity
density (e.g., \citealt{binggeli}).

With the advent of multiwavelength galaxy surveys, LFs began to be
constructed in the far-IR (e.g., \citealt{lawrence}), near-IR (e.g.,
\citealt{mobasher}), UV (e.g., \citealt{foca}), as well as for optical
emission line luminosities (e.g., \citealt{gallego}). It was generally
expected that these LFs will also follow the Schechter function, and
most of these studies, often times using LFs with very limited dynamic
range, confirmed such expectations. One striking exception was the LF
in the far-IR \citep{lawrence,saunders}, which was possible to
construct over a very wide dynamic range ($\sim 5$ dex in space
density) and which showed a significantly shallower decline at the
bright end than the exponential decline of the Schechter
function. While this difference between far-IR and other LFs have been
acknowledged (e.g., \citealt{buat98,takeuchi05}), the expectations set
by the perceived ubiquitousness of the Schechter distribution led some
to considered the far-IR LF as anomalous and perhaps deviating from
the Schechter form because of an AGN contamination (e.g.,
\citealt{bothwell}).

In this paper we show that there is a different explanation for such
deviations from the Schechter form.  We propose that there are two
fundamentally different galaxy distribution functions: (1) of the {\it
  stellar mass} (the mass function, MF\footnote{Since we will only be
  discussing the stellar mass, we will often be omitting the adjective
  ``stellar''.}) and of the (2) {\it star formation rate} (the SFR
function). Luminosity functions in different parts of the spectrum
will be related more to one type or the other.  Optical (especially in
bands past the 4000\AA\ break) and near-IR luminosities arise from
lower-mass stars that contain most of the stellar mass, therefore
these luminosity functions belong to mass function type. On the other
hand the emission in the UV, the nebular line emission (e.g., \ha,
[OII]) and the thermal IR is more closely related to young stellar
populations.  Therefore, such LFs, if properly dust corrected, should
belong to the SFR function type.

What are the true underlying functional forms of the stellar mass
function and the SFR function? Are they different? In the last decade
advances in stellar population modeling and the availability of large
surveys made the determination of the galaxy stellar masses possible
for a large number of galaxies. Being more fundamental than the
optical luminosity function, the MF received prompt attention. It too
was found to follow Schechter's functional form
\citep{cole01,bell03}. Indeed it can be said that the LFs in the
optical and the near-IR reflect the underlying Schechter-like
distribution of stellar masses.

Accurate SFR functions are more difficult to construct than the MFs
due to the caveats and larger uncertainties involved in deriving SFRs
(e.g., \citealt{kennicutt}). As pointed out, some LFs of SFR type were
found to be consistent with a Schechter form (UV and \ha) while
others, most notably the far-IR, was not. The current literature has
not fully explained this difference. The role of dust has been implied
in, e.g., \citealt{buat98,martin05,takeuchi05,reddy10}). A related
question, why LFs from young stars (UV and \ha) appear to have the
same Schechter-like distribution as LFs dominated by old stars
(optical and near-IR), received even less attention.

This study sets out to determine the intrinsic form of the SFR
function (in turn testing the adequacy of the Schechter function) and
to explain why different tracers produce different results. Knowing
the appropriate parametric form of the SFR function will help
interpret the observations at the range of redshifts and will
facilitate comparison with galaxy formation simulations.

To carry out the search for the functional form of the SFR function
(SFRF) we adopt a simple framework in which we produce volume-complete
mock samples of galaxies (\S 2.1) that are described by two
quantities: stellar mass and SFR, where the stellar mass is drawn from
a Schechter function, while the SFR is obtained by empirically
motivated relations between mass and SFR. We apply a series of three
such SFR--mass relations of increasing complexity, the final of which
being a relatively realistic representation of the observed SFR--mass
plane. We then study the SFR distributions produced by each relation
and discuss functional forms that can be used to describe them (\S
2.2--2.4).  Readers not interested in the details of these exercises
should skip to the summary in \S 2.5. We conclude that Schechter
formulation is not adequate for describing the SFR function. Instead,
functions that replace the exponential function at the high end with a
Gaussian represent a far better description. Next, in \S 3 we discuss
the implications of the use of non-Schechter functions for the
derivation of the SFR density. In \S 4 we apply the proposed
functional forms to the observed local SFRF and find excellent
agreement. In \S 5 we discuss the observed properties of LFs of SFR
type (UV, \ha\ and IR) and provide explanation as to why UV and \ha\
LFs appear to be well described using the standard Schechter function
despite being forms of a SFR function. We show that the Schechter-like
behavior is a coincidence stemming from the non-application of dust
corrections (or from application of only {\it average} statistical
corrections for dust) instead of dust corrections based on more robust
measurements of the attenuation on an {\it individual} galaxy
basis. On the other hand, we show that the non-Schechter form of
far-IR LFs more closely reflects the true SFR function and is not the
result of a purported AGN contamination.

Cosmological parameters $\Omega_m=0.3$, $\Omega_\Lambda=0.7$, $H_0=
70\, {\rm km\, s^{-1}\, Mpc^{-1}}$ are assumed throughout. We express
all stellar masses and SFRs assuming Chabrier IMF.

\section{SFR function} \label{sec:sfrf}

The goal of this paper to evaluate the adequacy of the Schechter
function for describing the star formation rate function and to
propose eventual alternatives to this function.  In order to perform
such an evaluation the ``true'' expected star formation rate function
must be known.  We derive the expected SFRF from the combination of
two relations which are well-determined locally: the stellar mass
function, and the stellar mass--SFR relation.

To produce SFR functions we construct mock samples in the following
way: we draw a large sample from a mass function that follows
Schechter parameterization. Then, to each mock galaxy we associate a
SFR based on a SFR--mass relation. We explore three different types of
stellar SFR--mass relations: (1) simple power-law relation between SFR
and mass with no scatter, (2) power-law relation with scatter, and the
(3) bimodal power-law relation with scatters in both modes. As
discussed below, each succeeding relation is meant to be more
realistic than the previous. The last should come very close to
describing true SFRs. For SFRFs resulting from these relations we test
the adequacy of the Schechter function and search for other functional
forms that potentially describe them better.

\subsection{Construction of mock samples}

To define the underlying Schechter mass function from which the mock
samples of galaxies are drawn we adopt parameters from \citet{panter}
who present a mass function based on the SDSS spectroscopic
sample. For our choice of Hubble constant and IMF these Schechter
parameters have the following values: mass function normalization
$\phi^{\star}=2.7\times 10^{-3}$ Mpc$^{-3}$ dex$^{-1}$, characteristic
mass (converted from Salpeter IMF by dividing the mass by 1.228, based
on \citealt{bc03} models) $\log M^{\star}_*=11.10$ (in solar mass
units, throughout), and the faint-end slope exponent of
$\alpha=-1.16$. This MF was constructed using data with $\log M_*
\gtrsim 7.7$, but we verify that its Schechter function fit is in
excellent agreement with the latest $6<\log M_*<8$ MF measurements
from \citet{baldry11}, which means that it can be safely extrapolated
to lower masses.

From this mass function we draw two volume-complete mock samples: (1)
a mass-limited sample with $M_*>10^8\, \ms$ and (2) a SFR-limited
sample with $\sfr >0.01\, \msperyr$. SFRs are assigned to each galaxy
according to one of the three SFR--mass relations as previously
described. Distinguishing between the two samples with different types
of limits is important because each affects the shape of the SFR
function differently. 

The exact choice of limits has no consequence on the inferences drawn
from the analysis, but we wish that they reflect some realistic
scenarios. For the mass-limited sample we take the limit to be $\log\,
M_*=8$, which is approximately the lowest mass for which statistically
large samples can be extracted from SDSS (e.g.,
\citealt{baldry08}). To this sample we do not impose any limits in
terms of SFR. For the SFR-limited sample we take the limit of 0.01
$\msperyr$, but no mass limit. Throughout this work SFRs represent true,
dust-corrected SFRs. This SFR limit matches the completeness limits of
surveys that target very nearby galaxies (such as the Local Volume
Legacy (LVL) survey, \citealt{kennicutt08,dale,lee}). Most of the
galaxies that produce stars at the rate around the limit are dwarfs
with $6<\log M_*<8$ \citep{johnson}. This SFR-limited sample will
probe a fraction of actively star-forming dwarfs at that mass, and
principally all star-forming galaxies above $\log M_*=8$ (\S\ 2.3). In
the local universe this SFR limit probes 99\% of the total SFR density
(\S\ \ref{sec:sfrd}).

The volume for the mock samples is $10^8$ Mpc$^3$ and was chosen to be
large enough so that the features of the SFR function are not
significantly affected by the Poisson noise over a large dynamic range
in space density ($\sim 5$ dex). The mass-limited sample contains
$3.4\times 10^6$ galaxies, and the SFR-limited sample up to $7.0\times
10^6$, depending on the SFR--mass relation used. Unlike real surveys
our samples are volume-complete by construction, so no completeness
corrections are required (for reviews of the construction of the
observed luminosity functions see
\citealt{johnston,takeuchi00,willmer}).

\subsection{Relation 1: SFR scales as a power of $M_*$ with no scatter}

Recent studies at low (e.g., \citealt{boselli,b04,s07}) and
intermediate redshifts (e.g., \citealt{noeske,elbaz}) have found that
for actively star-forming galaxies there is a relatively tight and
straight sequence in log $M_*$ vs.\ log SFR space, which has been
dubbed the {\it star-forming sequence} \citep{s07} or the {\it galaxy
  main sequence} \citep{noeske}. Straight sequence in log space is
equivalent to a power-law relation between SFR and mass ($\sfr\propto
M^{\beta}_*$). The reasons behind the existence of the relation are
currently the focus of many theoretical studies (e.g.,
\citealt{dutton,dave11}).

\begin{figure*}
\epsscale{0.7} \plotone{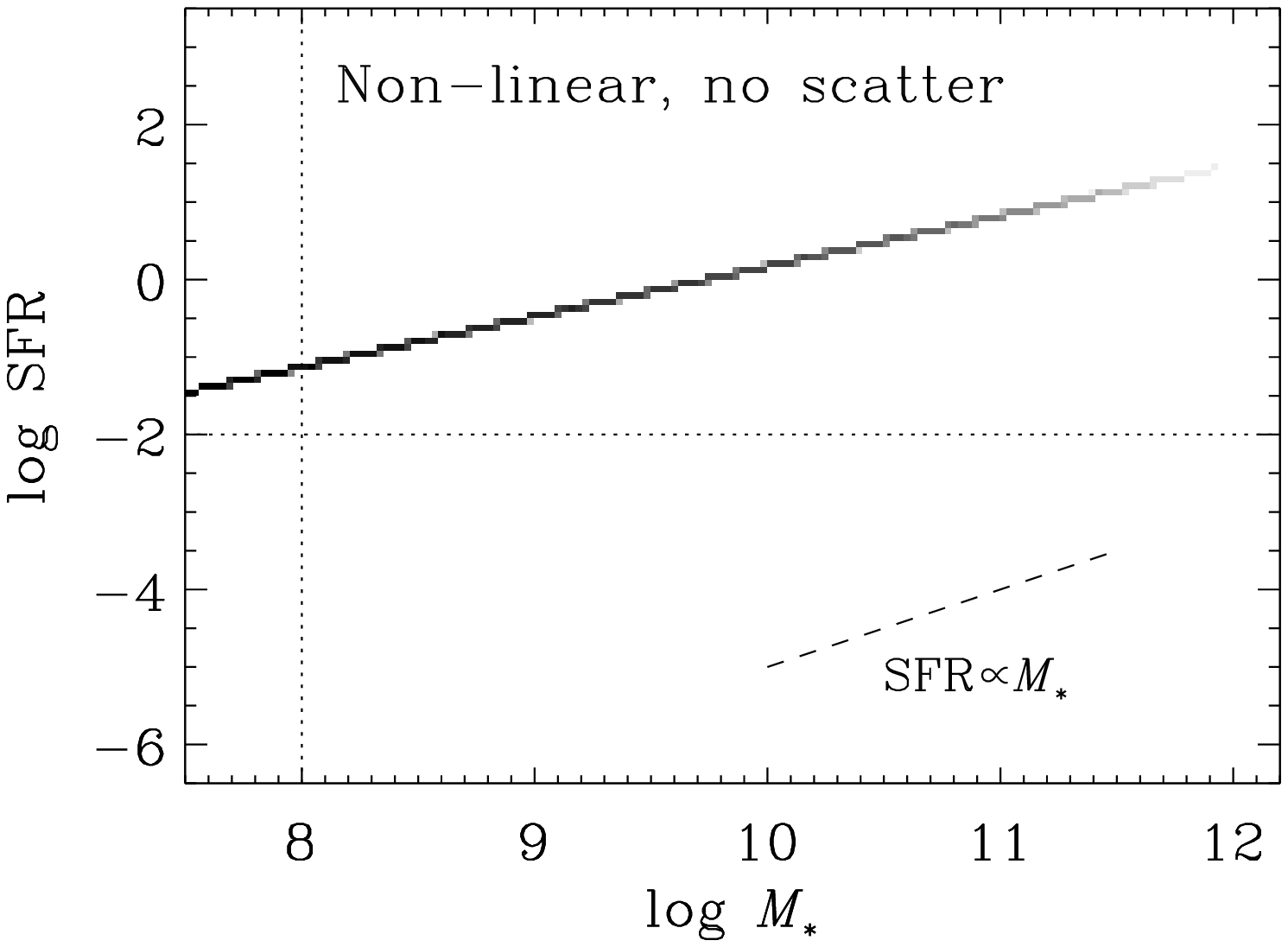}
\epsscale{1.2} \plotone{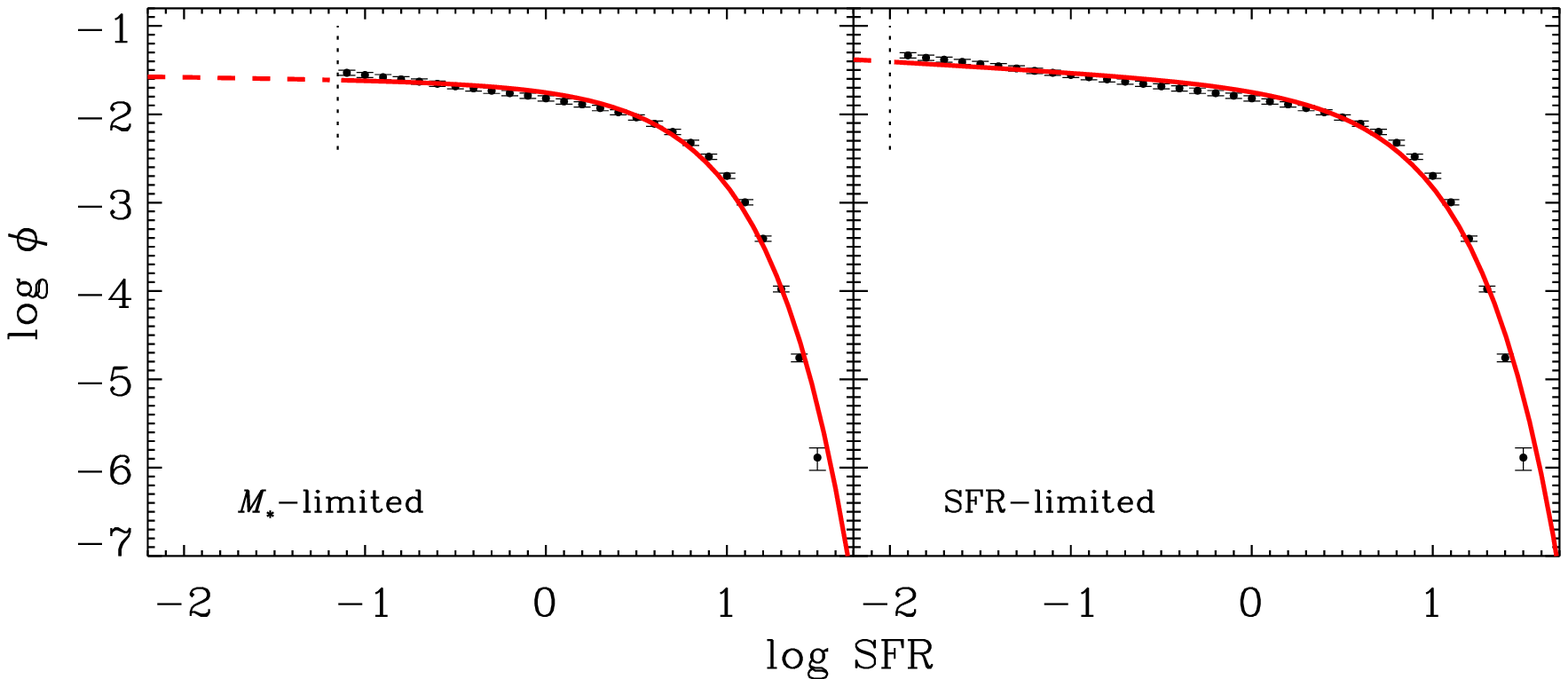}
\caption{Relation 1 and the resulting SFR functions. SFR scales as a
  sub-linear power of $M_*$, with no scatter, the simplest of the
  three relations that we explore. Upper panel shows the dependence of
  SFR on stellar mass. The relation appears jagged because it is
  represented as the bivariate density image, which is pixelated. The
  underlying stellar mass function is assumed to have the Schechter
  form. Dotted lines show limits for mass and SFR-limited
  samples. Lower panels show the resulting SFR functions for the
  mass-limited (left, $\log M_*>8$), and the SFR-limited samples
  (right, $\log \sfr>-2$). Vertical dashed lines indicate the lower
  cutoff used in the fitting. Red curves represent the best-fitting
  Schechter functions. Neither Schechter fit describes the SFRF
  accurately. \label{fig:noscat}}
\end{figure*}

Here we adopt the empirical relation derived in the local universe
($z\sim0.1$) from \citet{s07} (Eq.\ 11):

\begin{equation}
\log\, {\rm SFR} = 0.65\log\, M_*-6.33,
\label{eqn:sfr_mass}
\end{equation}

\noindent where masses and dust-corrected SFRs were obtained through
the use of UV/optical SED fitting of {\it GALEX} and SDSS fluxes of
galaxies falling in the star forming part of the BPT diagram
\citep{bpt}. The relation is shown in Figure \ref{fig:noscat} (upper
panel) and is sub-linear ($\beta=0.65$). Other studies have found
different values of $\beta$, but they are usually sub-linear
\citep{dutton,huang_s}.

Equation \ref{eqn:sfr_mass} is applied to masses drawn from the
Schechter mass function to obtain their corresponding SFRs. SFRs are
then binned in 0.1 dex intervals to obtain SFR functions shown in
Figure \ref{fig:noscat} (lower panels). Error bars represent Gaussian
approximation of the Poisson error and are typically extremely small
for most of the bins, meaning that the features of SFRFs are
accurately determined. To ensure more even weighting when performing
the fitting, we add in each bin 0.03 dex (7\%) of systematic error,
similar to errors in well-determined empirical LFs \citep{blanton}.

What functional form best describes the SFR functions
in Figures \ref{fig:noscat} (lower panels)? Their appearance suggests that
they would be well fit with a standard Schechter
function:\footnote{The logarithmic
  form of the standard Schechter function is expressed as:\\
  $\Phi_{\rm S}(\log X) d(\log X) =\ln (10)\, \phi^{\star}
  10^{(\alpha+1)(\log X-\log X^{\star})} \exp\left[-10^{(\log X-\log
      X^{\star})}\right] d(\log X)$.}

\begin{equation}
\Phi_{\rm S}(X) dX = \frac{\phi^{\star}}{X^{\star}} \left(\frac{X}{X^{\star}}\right)^{\alpha}
e^{-X/X^{\star}} dX,
\label{eqn:sch}
\end{equation}

\noindent where $X=\sfr$, $X^{\star}$ is the characteristic SFR,
$\phi^{\star}$ is the normalization (expressed in units of Mpc$^{-3}$
dex$^{-1}$ or Mpc$^{-3}$ mag$^{-1}$ throughout) and $\alpha$ is the
``faint''-end power-law exponent. We show the best fitting Schechter
function, obtained by minimizing $\chi^2$, as solid lines in Figures
\ref{fig:noscat} (lower panels).\footnote{Fitting is done in $\log
  \phi$.} The Schechter fits do not follow the SFRFs exactly. The
low-SFR (``faint'' end) slope of both fits tends to be shallower than
the SFRF points. Similarly, the knee of the fits appears to lie at
lower SFRs than what is expected visually. Finally, at the high end
the fits are slightly shallower than the SFR function. This mismatch
is corroborated with the large $\chi^2$ per degree of freedom (reduced
$\chi^2$) values of $\chi^2_r=6.2$ and 7.0 for the mass and
SFR-limited cases, respectively. Why is the Schechter function not a
perfect fit as may perhaps be expected?

The answer is that after the power-law transformation, the exponential
part of the Schechter function becomes modified into a S{\'e}rsic
function. Unlike the exponential function which has a fixed
high-end slope, the S{\'e}rsic function will have different slopes
based on the extra parameter that is featured in it. The reason why
the SFR function constructed in this way appears to be a
Schechter function is because on a logarithmic plot the {\it shapes}
of exponential and S{\'e}rsic functions are identical modulo the scale
factor, i.e., we can always pick an $x$ scale such that the two shapes
are exactly the same.

To properly fit the SFR distribution constructed using Eq.\
\ref{eqn:sfr_mass} the Schechter function needs to be modified by
introducing an additional parameter: the power-law exponent $\beta$
between the mass and SFR. We call such function the {\it extended
  Schechter function}:\footnote{The logarithmic expression for the
  extended Schechter is:\\ $\Phi_{\rm ES}(\log X) d(\log X)=\ln (10)\,
  \frac{\phi}{\beta}^{\star} 10^{(\alpha'+1)(\log X-\log X^{\star})}
  \exp\left[-10^{(\log X-\log X^{\star})/\beta}\right] d(\log X)$.}

\begin{equation}
  \Phi_{\rm ES}(X) dX = \frac{1}{\beta}\frac{\phi^{\star}}{X^{\star}} \left(\frac{X}{X^{\star}}\right)^{\alpha'}
  \exp\left[{-(X/X^{\star})^{1/\beta}}\right] dX,
\label{eqn:xsch}
\end{equation}

\noindent The exponential part of the standard Schechter function
became the S{\'e}rsic function, with $\beta$ being equivalent to the
S{\'e}rsic index. The extended Schechter function is the regular
Schechter function when $\beta=1$. We confirm that the extended
Schechter function fits the values of the SFR function perfectly and
retrieves parameters of the generating MF and SFR--mass
relation.\footnote{The low-end slope in the extended Schechter
  function is related to the low-end slope of the generating MF, which
  features in Eqn.\ \ref{eqn:sch} as $\alpha=\beta(\alpha'+1)-1$.} We
do not show these fits in Figure \ref{fig:noscat} (lower panels) since
they would simply pass through all the points with zero deviation.

The need for extending the Schechter function in order to model
certain distribution functions has recently been recognized in
\citet{bernardi10}, following \citet{sheth}, and in
\citet{hopkins10}. \citet{bernardi10} notice that galaxy sizes and
velocity dispersions scale as power laws with respect to the optical
luminosity ($\phi(L)$) and therefore remark that ``if $\phi(L)$ is
well fit by a Schechter function, it makes little physical or
statistical sense to fit the other observables with a Schechter
function as well.''\footnote{\citet{bernardi10} formulation of the
  extended Schechter function (their Eq.\ 9) appears to have an error
  (what is listed as $1/X$ should be $1/X^{\star}$). Furthermore,
  $\beta$ in \citet{bernardi10} and \citet{hopkins10} is the slope of
  mass vs.\ $X$ and therefore the inverse of our $\beta$.}
\citet{hopkins10} is the only work to our knowledge that has attempted
to apply the extended Schechter formulation to a SFR-like distribution
(a simulated IR LF).

The relation between mass and SFR assumed in this section is very
simplistic, so even the extended Schechter function will not
accurately reproduce all the features of real SFR distributions,
Nevertheless, it already demonstrates that SFR functions cannot be
adequately described by Schechter functions.

\subsection{Relation 2: SFR scales as a power of $M_*$, with scatter}
\label{ssec:scat}

\begin{figure*}
\epsscale{0.7} \plotone{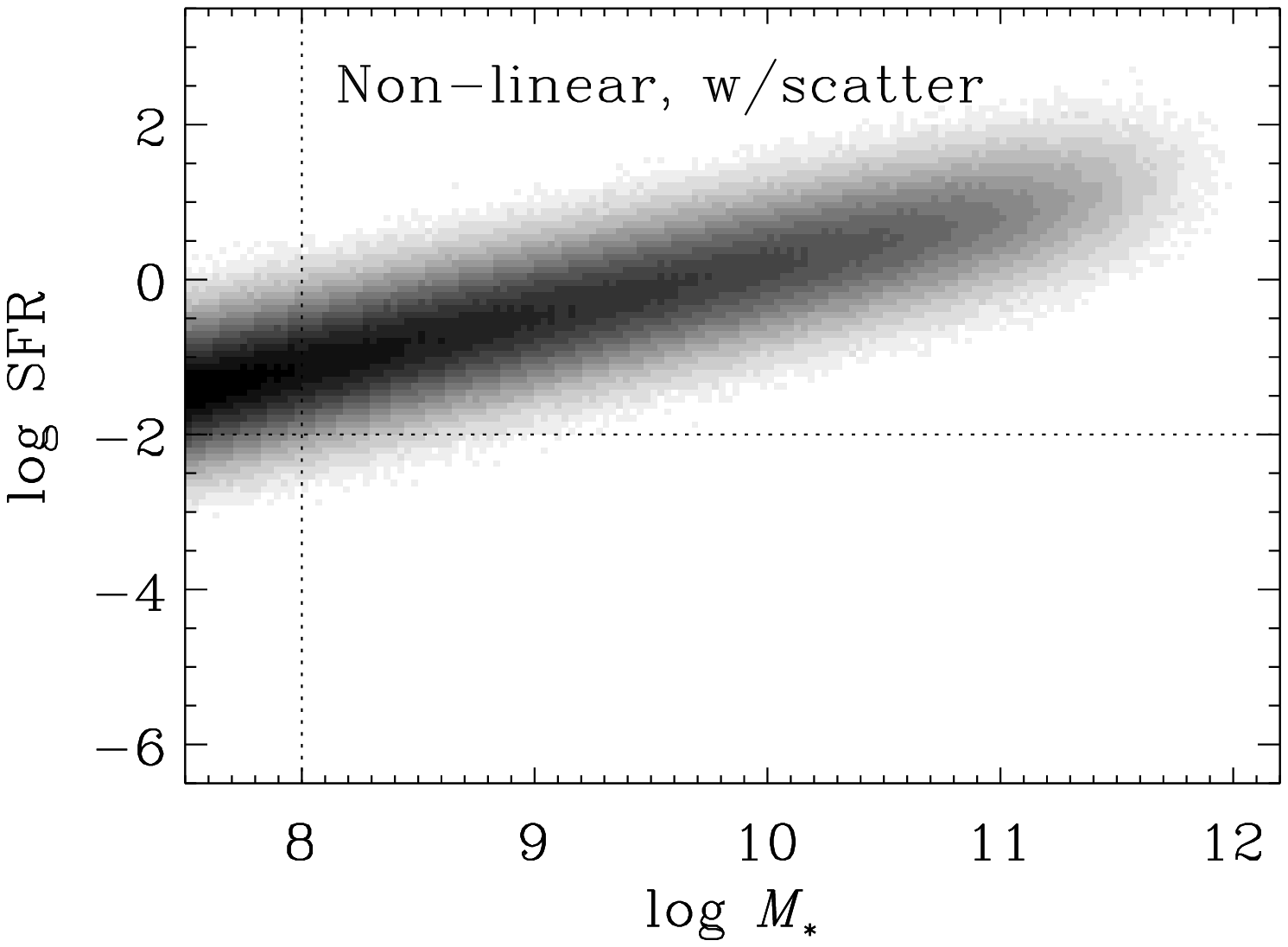}
\epsscale{1.2} \plotone{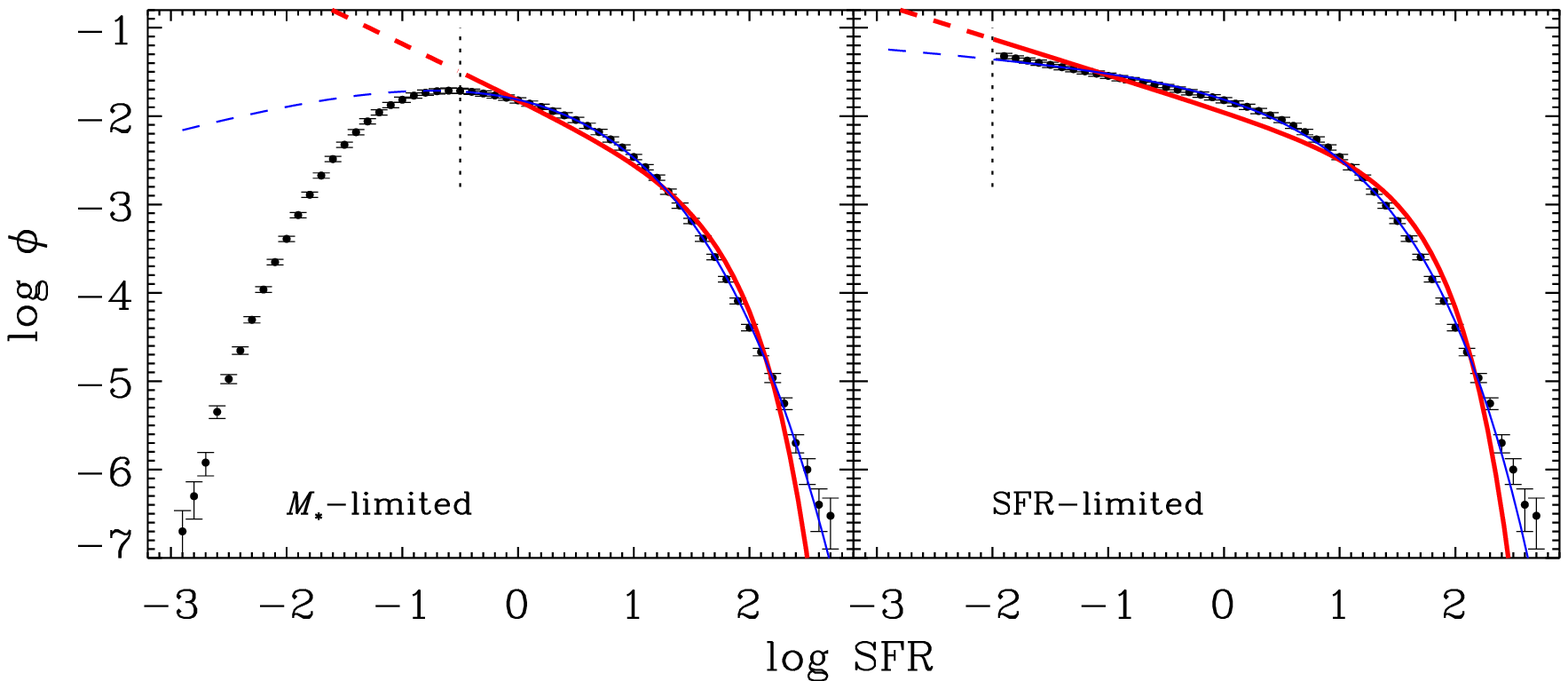}
\caption{Relation 2 and the resulting SFR functions. SFR scales as the
  power of $M_*$, with a Gaussian scatter of 0.4 dex in log SFR
  (log-normal in linear SFR). The upper panel shows the dependence of
  SFR on the stellar mass. Dotted lines show the limits of mass and SFR-limited
  samples. Lower panels show the resulting SFR functions for the
  mass-limited (left, $\log M_*>8$), and the SFR-limited samples
  (right, $\log \sfr>-2$). Vertical dashed lines indicate the lower
  limits used in fitting. Thick curves (red) represent the best-fitting
  Schechter functions, which describe the distributions very
  poorly. Thin curves (blue) are the best-fitting {\it extended} Schechter
  functions, which yield very good fits in the fitted
  regions. \label{fig:scat}}
\end{figure*}

While the SFR vs.\ mass relation is relatively tight, any scatter
around this relation that is not correlated with the mass would affect
the shape of the resulting SFR function. We model this scatter with a
Gaussian (in log SFR) of $\sigma=0.4$ dex, again based on the results
from \citet{s07}.\footnote{Throughout this paper we will refer to
  Gaussians, keeping in mind that in linear SFR these functions are
  actually log-normal.}  The scatter along the SF sequence increases
with mass from 0.3 to 0.4 dex, but the constant value is a reasonable
approximation for this exercise.  This scatter is predominantly
intrinsic, since the SFR errors in \citet{s07} for galaxies on the SF
sequence are $\approx 0.2$ dex. The SFR--mass relation with
scatter is shown in Figure \ref{fig:scat} (upper panel).

Figures \ref{fig:scat} (lower panels) show the SFR functions for the
mass and the SFR-limited samples. There are several notable
differences of these SFRFs with respect to the case with no
scatter. At the high end the tail is now much less steep extending to
higher SFRs (note that the horizontal scale in Figures \ref{fig:scat}
(lower panels) is much wider than in Figures \ref{fig:noscat} (lower
panels)), so the knee appears ``softer''.  The mass and SFR-limited
SFRFs start to differ significantly at $\log \sfr \lesssim -0.5$. The
SFR function of the mass-limited sample (Figure \ref{fig:scat} (lower
left panel)) now features a turnover and drops off at the low
end. This feature is due to the tail of galaxies close to the low-mass
limit that scatter below the mean SFR--$M_*$ relation. The fall off is
obviously a mathematical consequence of the presence of the mass
limit, but it is easy to confuse it with volume incompleteness, which
is why we show its effects separately. For the sample with no mass cut
(Figure \ref{fig:scat} (lower right panel)) there is no such turnover
and the SFRF continues to rise. Knowing how the sample is selected is
therefore very important in interpreting the ``faint'' end of any
observed SFR function. The fall off at the low SFRs can also be seen
in the cosmological simulations of the SFR function of \citet{dave11}
and the semi-analytic modeling of \citet{fontanot}, with both groups
using mass limits in their simulations.

In evaluating the functional forms that could be used to describe
these SFR functions we again start with the regular Schechter function
(Eq.\ \ref{eqn:sch}). For the mass-limited sample (Figure
\ref{fig:scat} (lower left panel)) we limit the fitting to the part
higher than the turnover, ($\log \sfr \geqslant -0.5$, dotted vertical
line), since obviously the Schechter function (or the extended
Schechter function) will not be able to reproduce the drop at low
values. The red line shows the best fit. One can see that the shape of
the Schechter function is quite inadequate, and the resulting
parameters are consequently of little value. The regular Schechter
function does not perform much better for the case of the SFR-limited
sample either (red line in Figure \ref{fig:scat} (lower right panel),
with both the low-end slope and the high-end drop too steep and the
knee too high. Formal reduced $\chi^2_r$ values are in both cases
extremely large (17 and 23, respectively). Note however that in some
real datasets where the data points have significantly larger error
bars and the dynamic range is small, a Schechter function fit could be
formally acceptable, leading one to believe that the distribution is
intrinsically of Schechter form.

On the other hand, the {\it extended} Schechter fit brings
significant improvements in describing both samples (blue lines in
Figures \ref{fig:scat} (lower panels)), with $\chi^2_r=0.3$ and 1.2
for the mass and SFR-limited distributions, respectively. Note,
however that even the {\it extended} Schechter function cannot
reproduce these SFRFs perfectly. The consequence is that the best-fit
parameters cannot be directly mapped back to parameters of the
underlying mass distribution and the relationship between the mass and
the SFR. In SFR--mass relation without scatter, $\beta$ represented the
slope of the SFR--mass relation. Now, the best fits have $\beta$
values of 2.08 and 3.03 for the mass and SFR limited distribution
respectively, in contrast with SFR--mass slope of 0.65.

The only way to reproduce these SFR functions exactly would be to
again reverse the process by which the SFRs were constructed. This can
be achieved with the extended Schechter function {\it convolved} with
a Gaussian.  Such a ``function'' would feature 5 or 6 parameters: four
of the extended Schechter function, the scatter $\sigma$, and also the
mass limit for mass-limited samples. We confirm that this function
fits SFRFs in Figures \ref{fig:scat} (lower panels) perfectly (fits
not shown), with the resulting parameters again having an
interpretable meaning. However, for any SFRF based on real data the
cost of two to three additional parameters with respect to the number
needed to describe the Schechter function will be too large -- the
resulting fits would suffer from a high degree of degeneracy and we
will consequently not consider this construct in further analysis.

\citet{bernardi10} provided an approximate analytical expression for
the effect of the measurement error on the extended Schechter function
in the limit of small $\sigma$ (their Eqn.\ 10 and 11).  However, that
form is not appropriate for the level of scatter encountered here,
which is dominated by large intrinsic scatter. This is exemplified by
the fact that the distribution corrected in that way does not preserve
the total number density of galaxies since the correction factor is
always greater than one.

\subsection{Relation 3: SFR is bimodal, each mode scales as a power of $M_*$, with scatter}

After including the scatter in the SFR--mass relationship we now add
one final element to bring mock SFRs close to the realistic
ones: galaxy bimodality. Galaxy bimodality is most often used to
describe the character of optical color distribution. The blue mode of
the color distribution corresponds to the star-forming sequence, which
is what was modeled in the previous sections. The optical red mode
corresponds to galaxies that do not belong to the star-forming
sequence and thus have little or no SF. We refer to them as passive
galaxies. They include optically red galaxies with SFRs measurably
different from zero (e.g., the green valley galaxies detected using
UV-optical colors; \citealt{martin,s07}) and the galaxies with upper
limits on SFR consistent with no SF.

If the passive galaxies were taken to have exactly zero SFR then the
modeling of the SFR distribution reverts to the unimodal case (\S
\ref{ssec:scat}), with the only difference being that the underlying
mass function would be for blue (star-forming) galaxies
alone. However, it is more realistic to characterize passive galaxies
with a range of non-zero SFRs, especially since passive galaxies on
the massive end can reach relatively high SFRs ($\lesssim 1 M_{\odot}
{\rm yr}^{-1}$; \citealt{cortese}).

To specify bimodal SFR--mass relations for use in our modeling we
again draw on the data derived in \citet{s07}. Expressions for
separate SF and passive sequences were not given in that paper, so we
determine them now by fitting in each mass bin two Gaussians in log
SFR. We find that the modeling of the SFR distribution at a given mass
with two Gaussians describes these distributions remarkably
well. Unlike in optical color where the colors of passive galaxies
quickly saturate (the red sequence), the passive sequence is quite
broad in log SFRs resulting in peaks that are not well separated, with
no pronounced dip between them.  The data probe the passive sequence
very well by reaching down to $\log (\sfr/M_*)=-14$. The peaks of
Gaussians yield the following SFR--mass relations that are very well
described with power laws. For the SF sequence:

\begin{equation}
\log\, {\rm SFR} = 0.54\log\, M_*-5.42,
\label{eqn:sfr_mass_sf}
\end{equation}

\noindent which is slightly shallower than the SF sequence defined
by SF galaxies selected using the BPT diagram (Eqn.\ \ref{eqn:sfr_mass}),

\noindent and for the passive sequence:

\begin{equation}
\log\, {\rm SFR} = 0.38\log\, M_*-5.20,
\label{eqn:sfr_mass_pass}
\end{equation}

\noindent the scatter of which varies from 1.5 dex at $\log M_*\sim 9$
to 0.7 dex at $\log M_*\sim 11.8$, and is again a combination of
measurement errors and, to a larger extent, the intrinsic scatter. For
simplicity, in our modeling we take the scatter of passive sequence to
be fixed at 1.1 dex, while for the SF sequence we use a scatter of 0.4
dex as previously.

At each mass we determine the passive fraction from the ratio of the
area below the Gaussian of the passive sequence and the total areas of
both Gaussians, which can be obtained from:

\begin{equation}
 f_{\rm pass}=\frac{N_{\rm pass}\sigma_{\rm pass}}{N_{\rm pass}\sigma_{\rm pass}+N_{\rm SF}\sigma_{\rm SF}},
\end{equation}

\noindent where $N$ is the height of the peaks.  It increases from
around 30\% at $\log M_*=8$ to 80\% at $\log M_*=11.5$. The passive
fraction is well described as a quadratic function of mass:

\begin{equation}
f_{\rm pass} =0.0534 \log^2M_*-0.905 \log M_*+4.144.
\label{eqn:fpass}
\end{equation}

\begin{figure*}
\epsscale{0.7} \plotone{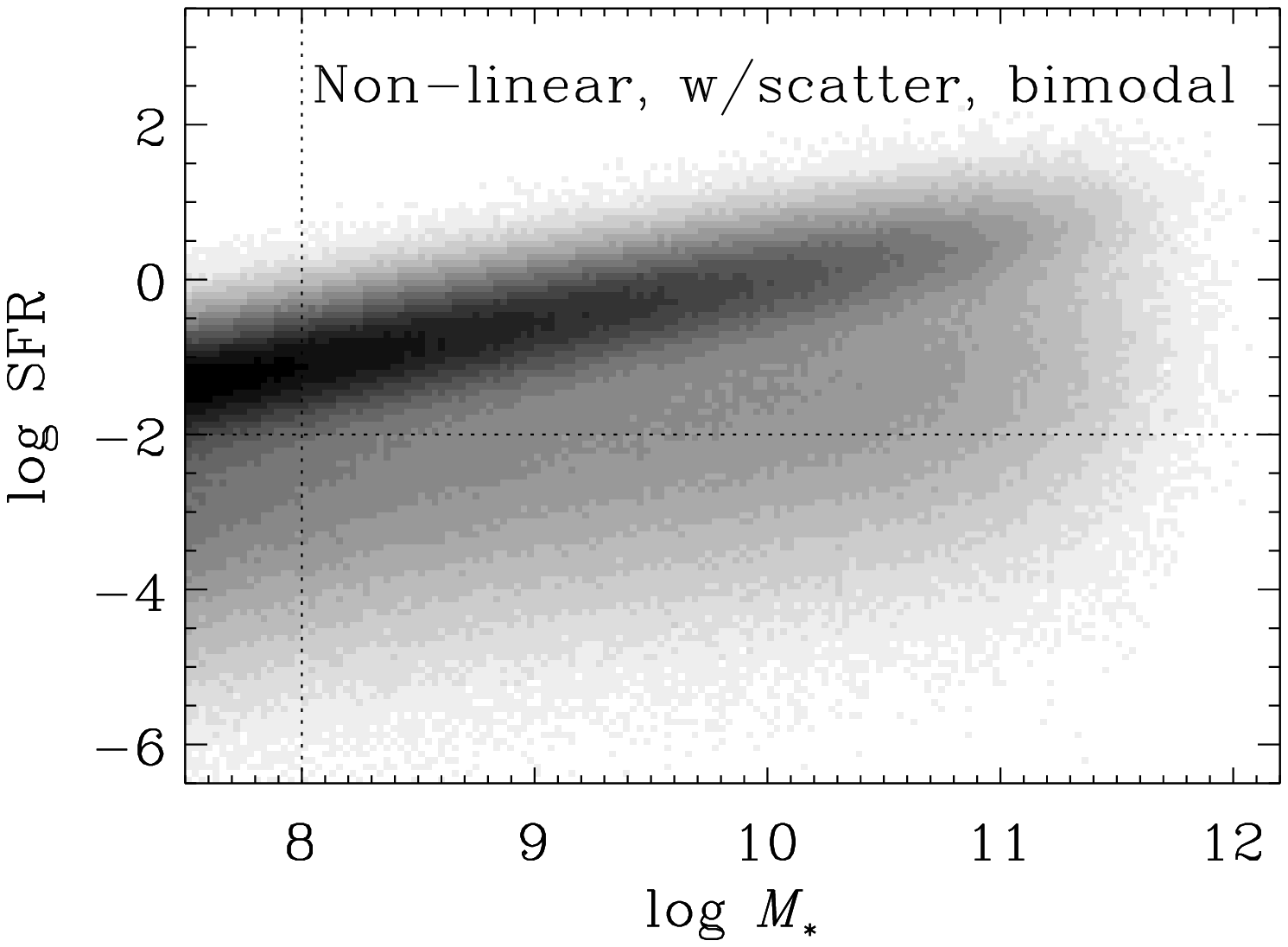}
\epsscale{1.2} \plotone{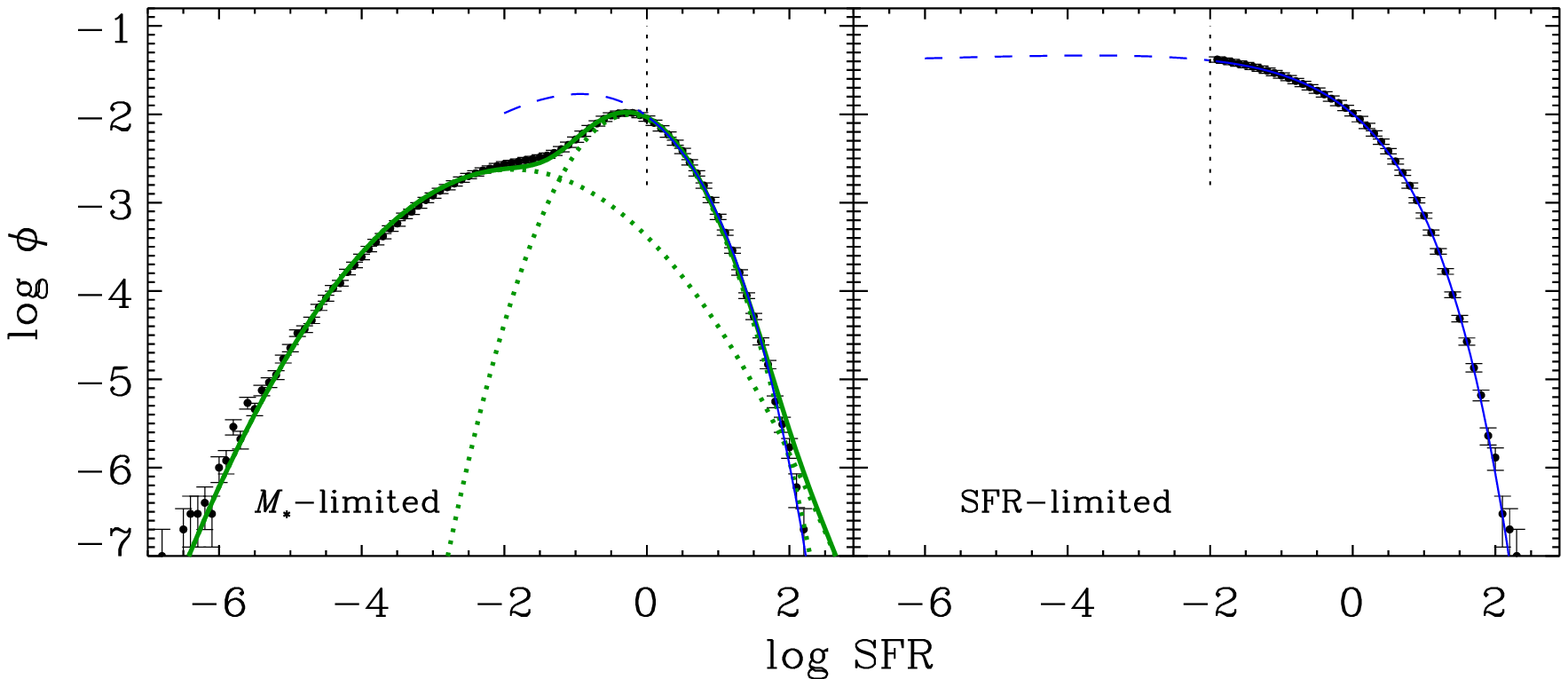}
\caption{Relation 3 and the resulting SFR functions. SFRs are bimodal
  and include the scatter around a star-forming and a passive
  sequence, with the fraction of galaxies in each sequence depending
  on mass as explained in \S\ 2.4. Upper panel shows the mock
  SFR--mass relation. SFRs on the passive sequence reach very low
  values. Dotted lines show limits of mass and SFR-limited
  samples. Lower panels show the resulting SFR functions for the
  mass-limited (left, $\log M_*>8$), and the SFR-limited case (right,
  $\log \sfr>-2$). Vertical dashed lines indicate the lower range used
  in the fitting the extended Schechter functions (thin, blue
  curves). Extended Schechter functions, as well as Saunders functions
  (not shown) yield good fits in the fitted regions. For the
  mass-limited case we also fit a double Gaussian (in log SFR, thick
  green curve), which produces a good fit for the entire SFRF.}
\label{fig:bim}
\end{figure*}

In Figure \ref{fig:bim} (upper panel) we show the SFR vs.\ mass values
of the simulated bimodal distribution. It was constructed so that for
each galaxy we first determine if it is passive or active using a
random number and Equation \ref{eqn:fpass}. Then we draw SFRs from
appropriate sequence and add the corresponding scatter. Since the
passive sequence has a fairly large scatter we do not allow a passive
galaxy to have a SFR greater than the SF sequence (Eqn.\
\ref{eqn:sfr_mass}) increased by $1 \sigma$ (i.e., 0.4 dex). In such
cases we draw a new value for the SFR.  For the mass-limited sample a
full span in mock SFRs is now $\approx 9$ orders of magnitude.

The resulting SFR functions are shown in Figure \ref{fig:bim} (lower
panels), for the mass and SFR-limited samples respectively. The two
distributions start to depart below $\log \sfr \approx 0$. The main
difference of the bimodal mass-limited SFRF compared to its unimodal
counterpart is an even more gradual drop to lower values due to the
relatively low SFRs of the low-mass passive galaxies, and the presence of the
inflection point (the ``dip'') at $\log \sfr \approx -1$, reflecting
the bimodal nature of the SFR distribution at each mass. On the other
hand, the SFR-limited distribution remains monotonic because at each
SFR it is dominated by galaxies in the star-forming sequence.

We again seek an appropriate functional fit for these SFR functions,
beginning with the SFR-limited distribution. Since the SFR-limited
distribution is very similar in the bimodal and unimodal case, the
regular Schechter function can be ruled out as a satisfactory form
based on previous considerations. As in the case of the unimodal
distribution, the {\it extended} Schechter function represents an
excellent fit ($\chi^2_r=0.3$). We show it in Figure \ref{fig:bim}
(lower right panel) with a blue line.

While the extended Schechter function represents an excellent
functional form for fitting the SFR-limited SFRF, recall that its
introduction was motivated by ``inverting'' the power-law mass--SFR
dependence in the simple unimodal case with no scatter. As a result of
that the exponential tail changed into a S{\'e}rsic function. Now, we
consider another alternative: a function that like Schechter and
extended Schechter maintains the power law at the low end, but now
features a Gaussian (in log SFR) at the high end. We
refer to it as the {\it Saunders function}, since it was first
proposed in \citet{saunders} to model LF at 60 $\mu$m. The linear form
of the Saunders function can be expressed as:

\begin{equation}
\Phi_{\rm S90}(X) dX = \frac{\phi^{\star}}{X^{\star}} \left(\frac{X}{X^{\star}}\right)^{\gamma}
\exp \left( -\frac{\log^{2} (1+X/X^{\star})}{2\sigma^2}\right) dX
\label{eqn:saund}
\end{equation}

\noindent Note that the ``Gaussian'' part of this function tends to a
Gaussian when $X>X^{\star}$, and to a constant when $X<X^{\star}$. This
modification (the addition of 1 in the argument of log) allows the
low end to transition smoothly into a power law.

Fitting the Saunders function to a SFRF constructed using the bimodal
SFR--mass relation and SFR-limited sample (Figure \ref{fig:bim} (lower
right)) we obtain a very good fit with $\chi^2= 0.5$. This
is slightly worse than what the extended Schechter fit yielded
($\chi^2=0.3$), however, whether one or the other is better will
depend on the level of scatter, especially in the SF sequence. We know
that for the unimodal SFRs, in the limit of zero scatter the extended
Schechter function is a perfect analytic description, but this will be
less true for large scatter. For the scatter assumed here (0.4 dex for
the SF sequence), both functions are basically equally good
approximations. 

The Saunders function has one significant advantage over the extended
Schechter function---its parameters are less easily perturbed by
errors in SFRF and, therefore, the relative accuracy of the parameters
corresponding to the Saunders fit is higher. In the case of the
bimodal SFRF discussed here the Saunders fit has $\sim2.5\times$
smaller errors in log SFR$^{\star}$ and log $\phi^{\star}$ and
$5\times$ smaller uncertainty in the ``faint''-end slope. This is the
result of the significantly lower level of covariance among the
parameters of the Saunders function. On average the Pearson
correlation index for Saunders function parameters is 0.70, while it
is 0.94 for extended Schechter. Most importantly, the parameters
describing the ``faint'' and the ``bright'' ends have a correlation of
only 0.33 in Saunders function ($\gamma$ and $\sigma$), and yet 0.87
in the extended Schechter function ($\alpha'$ and $\beta$).

While the Saunders function has been proposed for describing the 60 $\mu$m
LF, which can be considered a type of SFRF, the motivation for its
introduction was simply to provide a functional form that better
describes the real data than the standard Schechter function. Here we
show that the reasons behind such good representation have to do with
the nature of SFR and its relation to the stellar mass.
 
Turning now to the mass-limited SFRF (Figure \ref{fig:bim} (lower left
panel)), if we only aim to fit the part to the right of the turnover
($\log \sfr \geqslant 0$), then the same conclusions hold as in the
SFR-limited case: the extended Schechter function represents an
excellent fit ($\chi^2_r=0.5$; blue line in Figure \ref{fig:bim}
(lower left panel)). Similarly well does the Saunders function (not
shown). Is it possible to successfully fit the entire distribution
including the low-end drop and the dip? Both the extended Schechter
function and the Saunders function are monotonic and feature a power
law at the low-end, so they will be incapable to reproduce the
inflection or the low-end drop. Thus we test a new functional form: a
{\it a double (i.e., composite) Gaussian} function (in log SFR), in
which each Gaussian should fit one of the two modes of SFR
distribution. This function is given in the log form as:

\begin{equation}
\Phi_{\rm GG}(\log X) d(\log X) = (\phi^{\star}{_P}G_{P} + \phi^{\star}{_S}G_{S})  d(\log X), 
\end{equation}

\noindent where

\begin{equation}
G_{P} = \frac{1}{\sigma_{P}\sqrt{2 \pi}}\exp \left( -\frac{\log^{2} (X/X^{\star}_{P})}{2\sigma_{P}^2}\right),
\end{equation}

\noindent represents the Gaussian corresponding to the passive
population and

\begin{equation}
G_{S} = \frac{1}{\sigma_{S}\sqrt{2 \pi}}\exp \left( -\frac{\log^{2} (X/X^{\star}_{S})}{2\sigma_{S}^2}\right),
\end{equation}

\noindent the Gaussian of the SF population, each with its own
standard deviation and peak position. Double Gaussian features six
parameters, but the covariances are weak between each set of three. 

Indeed, when we fit the double Gaussian we obtain very good results
(solid green line in Figure \ref{fig:bim} (lower left panel), with the
individual components shown with dotted green lines; on a log-log plot
the double Gaussian is represented as two parabolas). The fitted
Gaussian standard deviations are $\sigma_{P}=1.0$ and
$\sigma_{S}=0.5$, reflecting the scatters of the passive and SF
sequences. As in the case of the Saunders function, the parameters of
the double Gaussian are significantly less sensitive to SFRF
uncertainties than the parameters of the extended Schechter
function. While much better than any other practical alternative, the
double Gaussian is not a perfect fit ($\chi^2_r=1.9$). It produces
slightly stronger inflection (deeper dip) than what is seen in the
SFRF. Also, recall that when we constructed SFRs we required that SFRs
from the passive sequence do not exceed SFRs of the SF sequence by
more than 0.4 dex. However, no such restriction is imposed in the
fitting, so the passive Gaussian extends a bit too much at
the high end. Clipping the passive Gaussian where it starts to exceed
the SF Gaussian brings the reduced $\chi^2_r$ to 1.5.

Note that the success of the double Gaussian in fitting the SFRF is
not a mere consequence of the fact that in each mass bin we modeled
SFR--mass relations as the sum of two Gaussians because the slope of
the SFR--mass relation is significantly larger than zero (i.e., the
Gaussians from different mass bins have different centers in SFR).

To our knowledge the double Gaussian was not previously considered as
a functional form for the SFR function. A single Gaussian (lognormal
function in linear SFR) has been suggested by
\citet{martin05}. However, such form (a parabola on the log-log plot)
is apparently inadequate when the details of the SFR function are
considered, i.e., when it is measured precisely below log SFR$\lesssim
0$. \citet{fontanot} notice the ``double peak'' feature in their
$0.4<z<1.8$ mass-limited SFR functions and suggest that this
``peculiar feature'' might be connected to bimodality. Our analysis
shows that bimodality is indeed the explanation.

\subsection{Summary: functional forms describing SFR function}

Summarizing the modeling section we conclude that the Schechter
function is an inadequate description for any realistic SFR
function. Instead, SFRFs derived from the SFR-limited sample can be
sufficiently well modeled using either the extended Schechter function
or the Saunders function (a power law with a Gaussian decline at the
high end). The latter is recommended because its parameters, being
less covariant, will be more stable and yield higher relative
accuracy. For the mass-limited SFRFs the double Gaussian represents an
excellent solution. Note that in all of these cases the fitted
parameters will not directly be interpretable as the parameters of the
underlying mass function or the SFR--mass relation, but they will
provide robust descriptions of SFRFs that can be compared from one
study to another.

\section{Star formation rate density} \label{sec:sfrd}

There are two primary reasons for which describing the SFR function
with an analytic function form is useful. One is to characterize this
distribution through parameterization, to facilitate the study of its
evolution with redshift. The second is to use the parameterization to
infer, by integration that involves extrapolation, the total SFR {\it
  density}. In that case the Schechter formulation is especially
practical because its cumulative distribution function is finite when
$X \to 0$ and has a simple analytic expression. Since, as we have
demonstrated, the Schechter formulation does not provide an adequate
description of the observed SFR function, alternative methods are
needed for inferring the total SFR density. While all of the
alternative functions that we have considered (the extended Schechter,
Saunders and the double Gaussian) also have finite cumulative
distributions, they cannot be integrated to yield functions in a
closed form. Therefore, we will instead employ our bimodal model
(constructed based on $z \sim 0.1$ SFRs) to provide numerical
correction coefficients to be applied to the SFR density
determinations in the local universe obtained from the direct
numerical integration of the observed SFRF down to some SFR limit,
based on sample with some mass limit.

\begin{figure}
\epsscale{1.2} \plotone{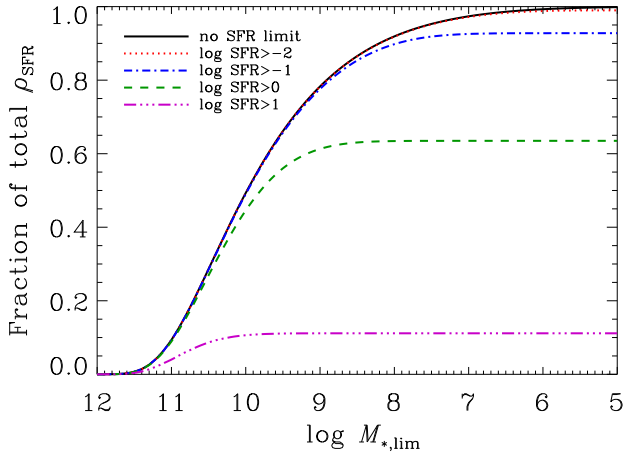}
\caption{Fraction of the total SFR density that is accounted for by
  integrating mock bimodal SFRF down to a given mass limit ($x$ axis), and
  SFR limit (different curves). To probe 90\% of the SFR density in
  the local universe requires sampling galaxies to $\log M_*\approx 8$
  and $\log \sfr >-1$. The figure can be used to determine the
  correction factors (1/fraction) to be applied to the SFR density
  determinations in the local universe ($z\sim0.1$) obtained from the
  direct numerical integration of the SFRF, and to estimate the yield
  of planned SFR surveys.}
\label{fig:sfrd_lim}
\end{figure}

The fraction of the total SFR density that will be present in the
observed local SFRF to a given limit is shown in Figure
\ref{fig:sfrd_lim}. The mass limit can be read continuously from the
$x$-axis, while each of the curves represents some SFR limit. The
black solid curve shows the fraction accounted for at different mass
limits if there was no SFR limit. Going to $\log M_*=8$ in such case
(corresponding to the mass limit that we explored heretofore) probes
92\% of the local SFR density; i.e., the values obtained from
integrating the SFRF in Figure \ref{fig:bim} (lower left panel) would have to be
corrected upward by 9\%. It should be noted that in the presence of a mass
limit the correction is not due to mere extrapolation of the SFRF, but
serves to correct for lower mass galaxies that contribute at a range
of SFRs. Having the SFR limit of $\log \sfr=-2$ (red dotted curve)
leads to negligible difference with respect to the case with no SFR
limit. On the other hand, a limit of $\log \sfr=0$ (green dashed curve)
captures only 64\% of the SFR density in the local universe.

We emphasize that if only the SFR limit is present one could recover
the missing SFRD by simply extrapolating the faint-end power law tail
(of course, assuming the slope is well constrained), but if the mass
limit is also present one would have to account for it too, using,
for example, the provided figure or performing modeling similar to
that presented here. Similar technique can also be used to estimate the
yield of a planned survey or to estimate corrections due to
incompleteness.

\section{The observed local SFR function}

\begin{figure}
\epsscale{2.5} \plottwo{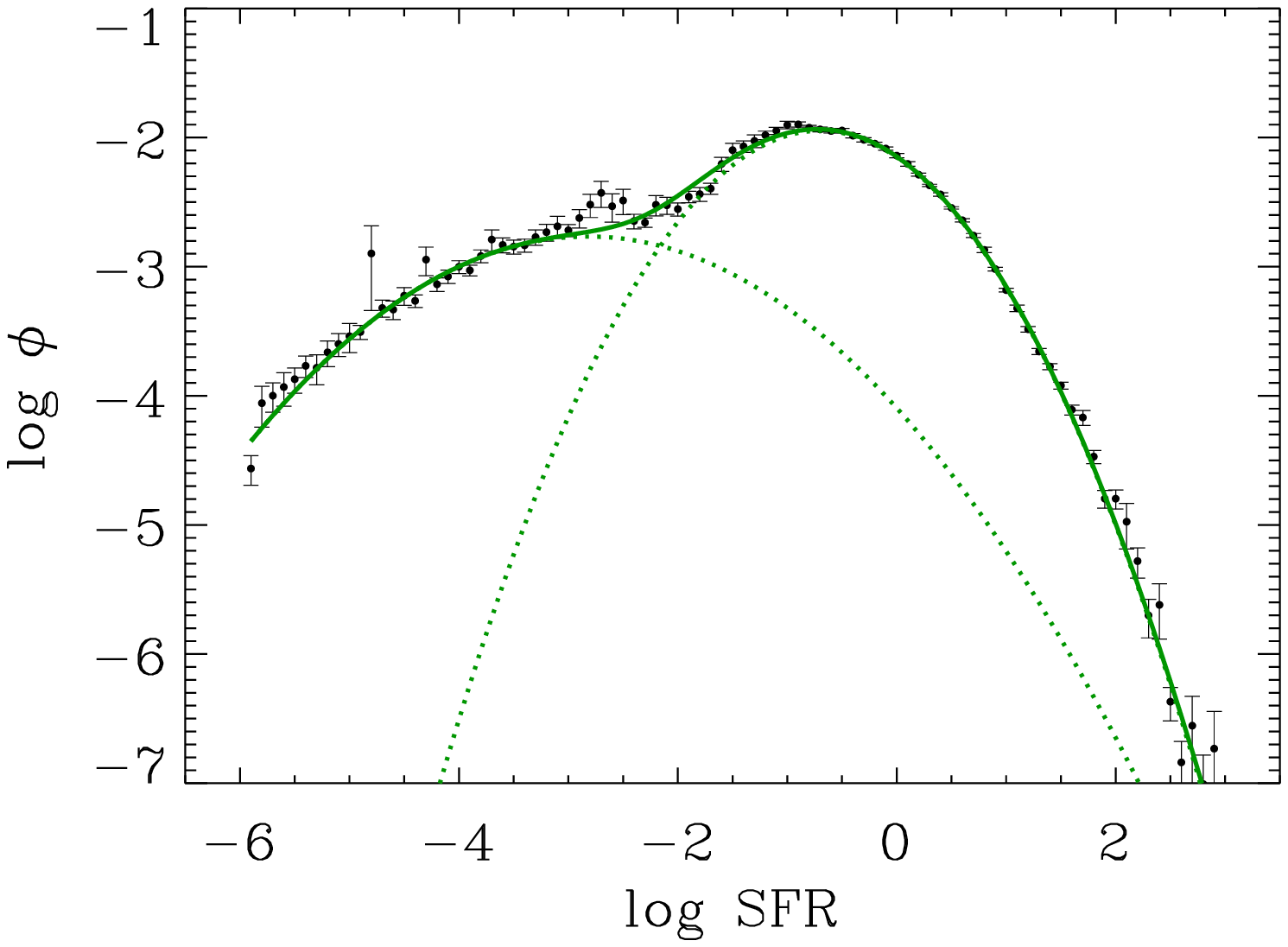}{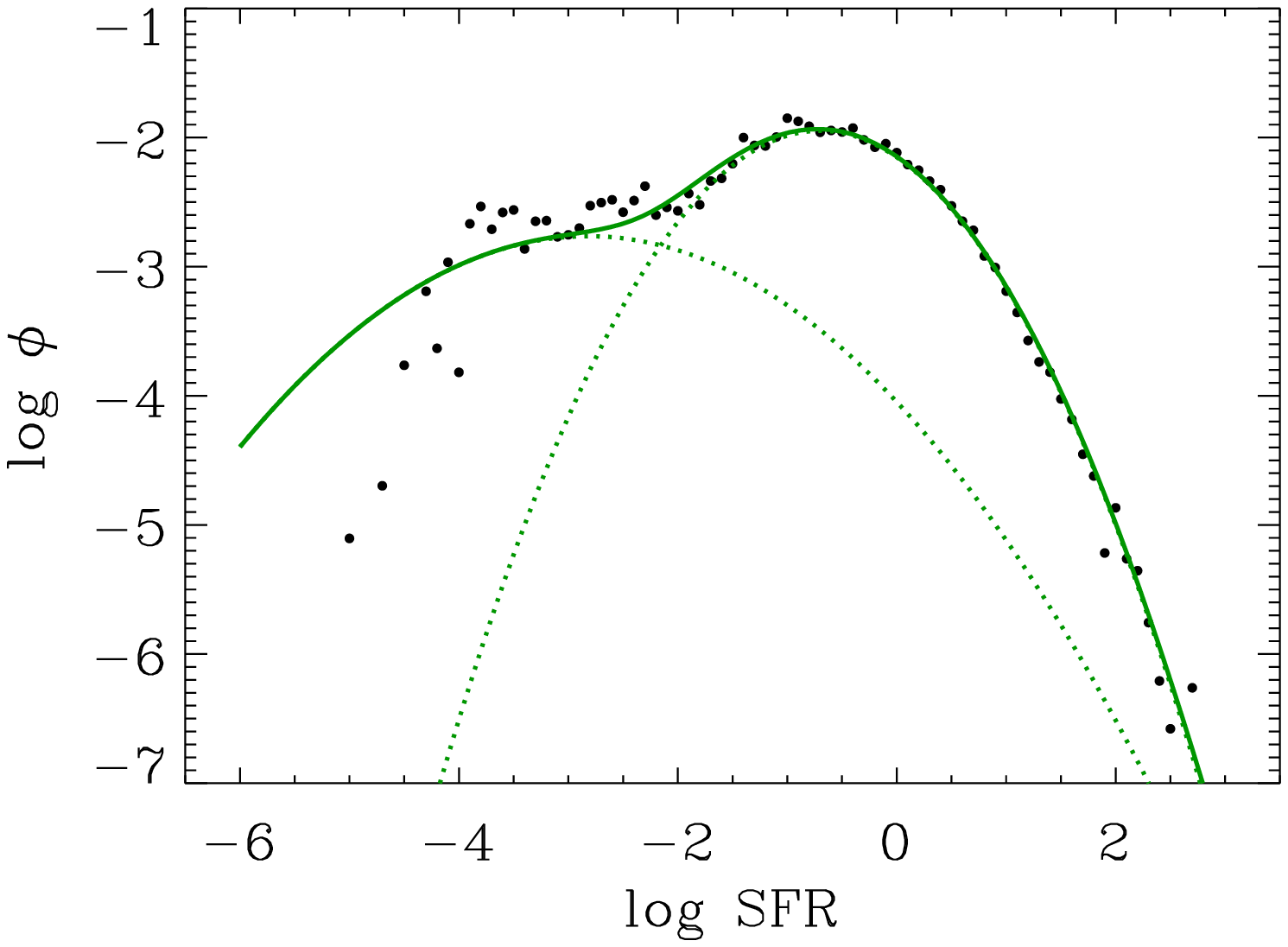}
\caption{Upper panel. The observed $z\sim 0.1$ SFR function obtained
  from \citet{s07} data where SFRs where obtained from UV and optical
  SED fitting of an optically selected sample with $\log M_*>8$. SFRF
  is a composite of each galaxy's SFR probability distribution, not a
  single value. Error bars are determined from the standard deviation
  of bootstrap samples. Green curves represent double Gaussian
  fits. Lower panel. Same as Figure above except that the SFRF was
  constructed such that each galaxy's SFR was given by a single value
  (mean of the probability distribution). Green curves are repeated
  from the upper panel. Most of the SFRF stays unchanged.}
\label{fig:obs}
\end{figure}

Next we investigate the shape of the {\it observed} SFR functions and
test whether they can indeed be described with the functional forms
determined based on simulated SFRs in \S\ \ref{sec:sfrf}. We construct
the SFR function from UV/optical-based SFRs of $\approx 50,000$
galaxies from \citet{s07}. This dataset only has a mass limit ($\log
M_*=8$) and can formally yield specific SFRs as low as $\log
(\sfr/M_*)=-12$, which corresponds to very low SFRs even for massive
galaxies. The SFRF from constructed from these data is shown in Figure
\ref{fig:obs} (upper panel). As expected for the mass-limited sample,
the SFRF drops off at low values. Overall, its shape is quite similar
to the bimodal {\it mock} SFRF presented in Figure \ref{fig:bim}
(lower left panel). This is not too surprising because the
construction of the mock SFRs was guided by these observations, but it
should be kept in mind that the SFR--mass relations we used
incorporated some simplifying assumptions (e.g., fixed scatter). We
previously concluded that the {\it double Gaussian} function is the most
suitable description of the mock bimodal SFRF with a mass limit. We
fit that function to the observed data points and indeed obtain a very
good fit (shown by green curves in Figure \ref{fig:obs} (upper
panel)). The resulting goodness of fit is $\chi^2_r=1.7$. The
parameters of the best fit are:

\begin{equation}
  \log \sfr^{\star}_P=-2.82, \quad \sigma_P=1.14, \quad
  \log \phi^{\star}_P=-2.31
\end{equation}
\begin{equation}
  \log \sfr^{\star}_S=-0.70, \quad \sigma_S=0.72, \quad
  \log \phi^{\star}_P=-1.69.
\end{equation}

\noindent To construct the SFRF we co-added $1/V{\rm max}$-weighted
{\it probability distribution functions} for each galaxy's SFR. This
is possible because the SFRs in \citet{s07} were obtained using the
Bayesian SED fitting. In Bayesian SED fitting each model SED
contributes (proportionally to $e^{-\chi^2/2}$) to the probability
distribution of a galaxy parameter (such as the SFR). Using full
probability distributions for each galaxy's SFR (instead of a singular
value) has the advantage that it produces more realistic distributions
for ensemble of galaxies. Since full probability distributions of SFRs
are not always available, we also considered the SFRF where each
galaxy is represented by a single value of SFR (the mean of the
probability distribution). Such SFRF is shown in Figure \ref{fig:obs}
(lower panel). We overlay it with the best double Gaussian fit
obtained using full probability distributions. There is an excellent
agreement between the two especially in the region fit by the
star-forming Gaussian, including the high-end tail. The latter means
that the distribution in the high-end tail is not due to some galaxies
having broad SFR probability distributions reaching very high
values. Discrepancies start to appear only below $\log \sfr \lesssim
-3$, because galaxies with very low SFRs usually have broad (poorly
constrained) probability distributions, so when these are collapsed
into a single value for SFR their extent towards very low values gets
somewhat compressed. In any case, such low SFRs have very little
effect on any global characterization of the SF.

Our mock SFR functions in \S\ 2 assumed that masses are drawn from a
Schechter function. Deviations of the MF from the Schechter form are
known \citep{baldry08}, but they are relatively small and mostly
pertain to lower-mass passive galaxies \citep{peng10}. The
ability to reproduce the observed SFRF using functions suggested by
mock SFRFs confirms that these deviations are not significant in the
context of this study.

\section{Discussion}

Our analysis of mock SFRFs and observations from \citet{s07} show that
real SFRFs have significant departures with respect to a Schechter
function. One usually determines the SFR based on the luminosity of a
tracer population of young stars, such as the far-UV continuum
luminosity, emission line luminosity (e.g., \ha, OII or Pa$\alpha$),
or from the dust luminosity in some part of the IR SED (PAH lines,
mid-IR continuum, far-IR continuum or the total IR luminosity). It is
then to be expected that the {\it luminosity functions} of these
various tracers would show similar departures from the Schechter
distribution as the SFRF. While such departures have been known for a
long time in the IR (especially the far-IR; e.g.,
\citealt{lawrence,saunders,takeuchi}), there is a general consensus
that the far-UV LF as well as the \ha\ LF {\it do} to a large degree
agree with the Schechter function (e.g., \citealt{wyder} for UV and
\citealt{gallego,ly11} for \ha). How can we explain this apparent
inconsistency? As the analysis in this section will show, the primary
reason for this is because the observed, {\it uncorrected} UV and \ha\
LFs have a form that is similar to a Schechter function by coincidence
(similarly for LFs where dust is ``corrected'' by applying average
statistical relations). On the other hand LFs, where the luminosity of
each galaxy is individually dust corrected (thus becoming a true SFR),
do show large departures from Schechter form in line with our analysis
for SFRFs. The departure of the LFs from the Schechter form, even with
individually applied dust corrections, is more difficult to recognize
in high-redshift studies which feature small samples with the
resulting LFs having limited dynamic ranges, which is why this section
will discuss more robust observational evidence from lower redshifts.

\subsection{UV LFs}

\begin{figure}
\epsscale{2.5} \plottwo{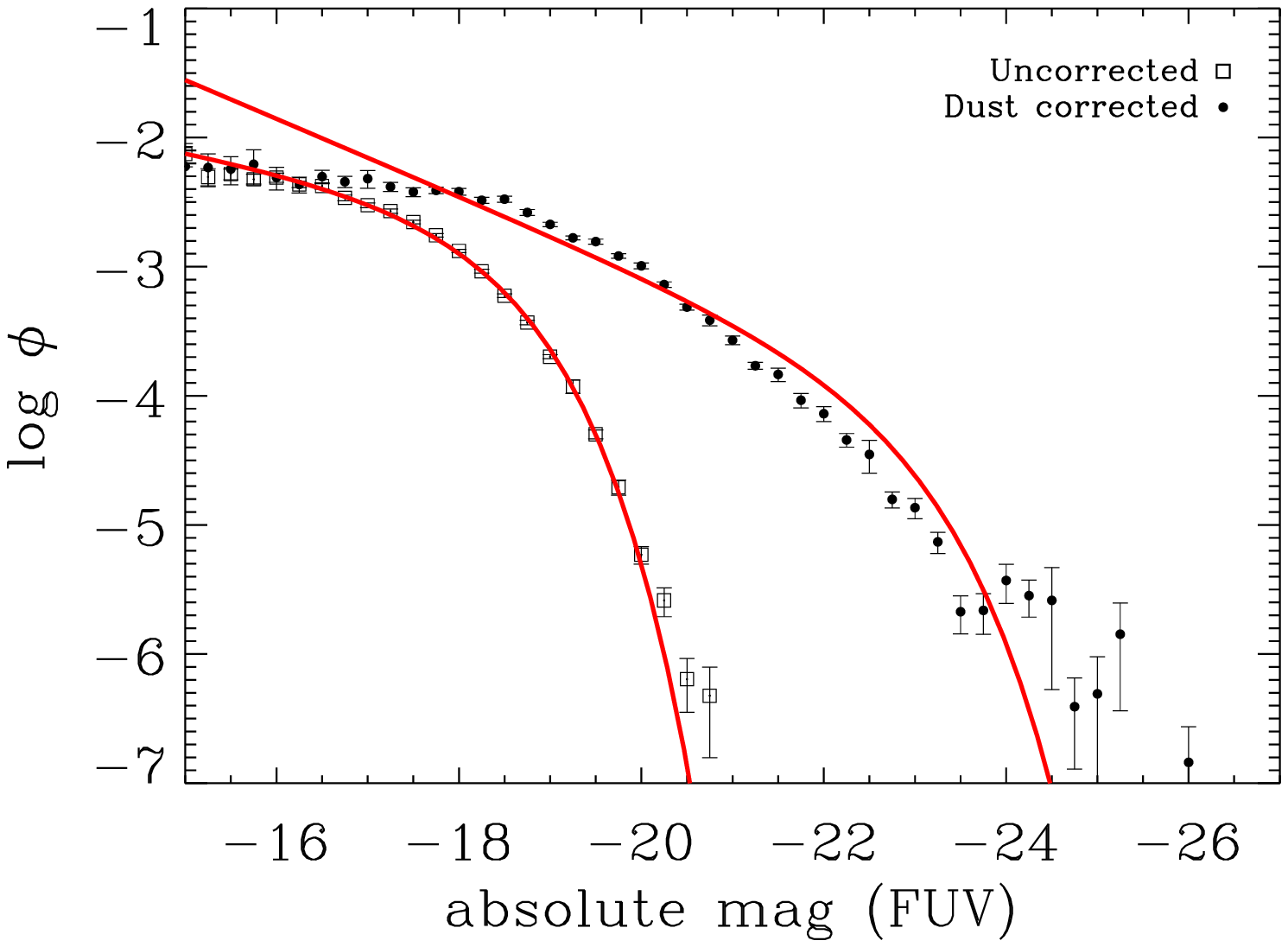}{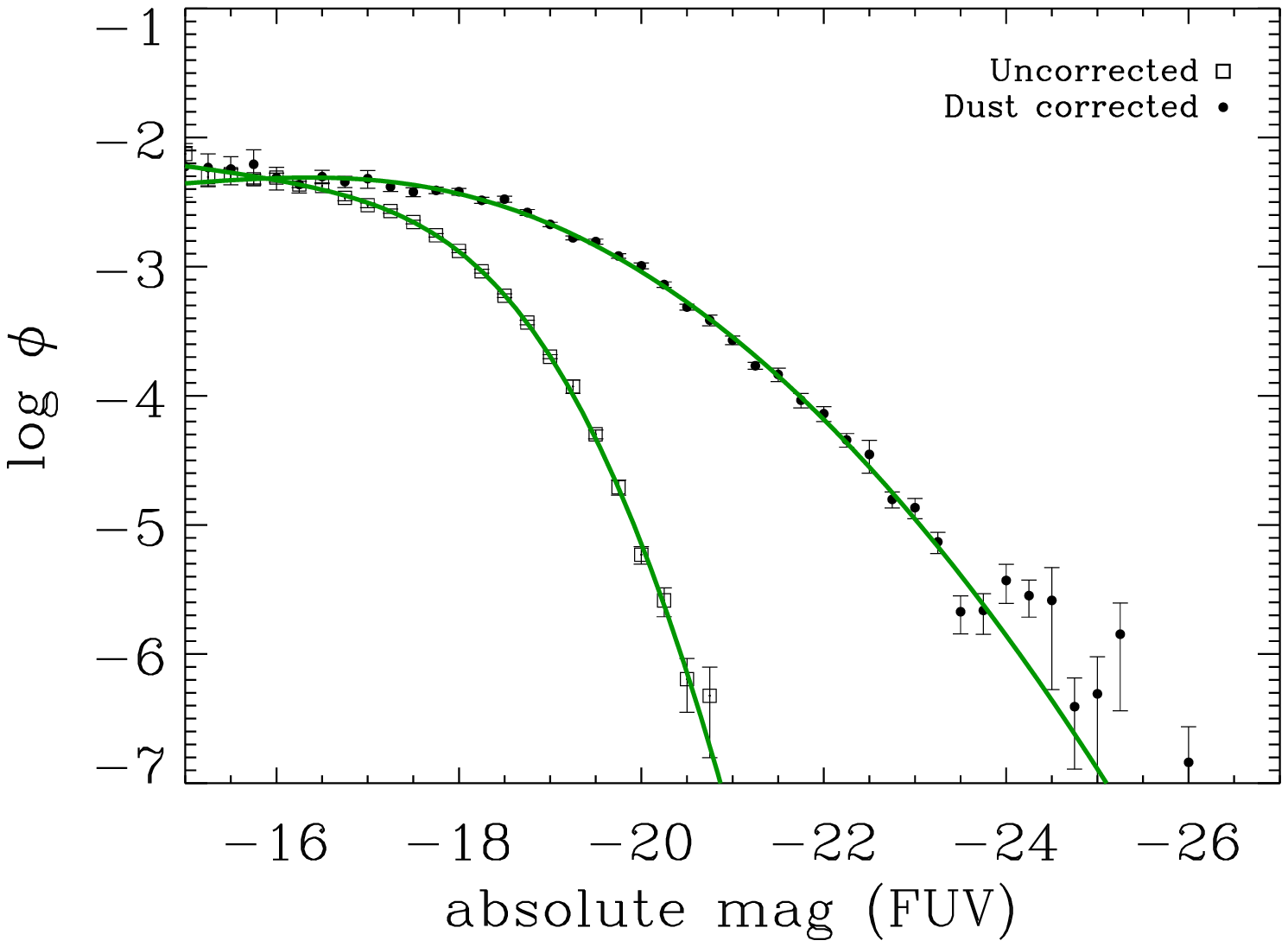}
\caption{Upper panel. Far-UV luminosity functions based on \citet{s07}
  sample with (dots) and without (open squares) dust correction. Dust
  corrections are determined on galaxy-by-galaxy basis. Error bars are
  determined from the standard deviation of bootstrap samples. Red
  lines represent best-fitting Schechter functions. Schechter function
  is an acceptable fit only for the uncorrected LF.  Lower panel. LFs
  are repeated from upper panel, but the fitting functions (green
  curves) are now \citet{saunders} functions which feature Gaussian
  high ends instead of exponentials in the Schechter function. They
  provide excellent fits for both the uncorrected and the corrected
  LFs. A LF corrected using average statistical relations (e.g.,
  between dust attenuation and stellar mass) would also yield steep
  Schechter-like bright-end slopes, and not the much shallower slope
  of the true SFRF.}
\label{fig:fuv_lf}
\end{figure}

The characterization of galaxies in the ultraviolet has greatly
improved with the launch of {\it GALEX}, which surveyed most of the
sky in two UV bands (FUV, 1500 \AA; NUV, 2300 \AA). Based on the {\it
  GALEX}/2dF observations of $\sim 1000$ galaxies \citet{wyder}
presented FUV and NUV LFs for local ($z<0.1$) galaxies. These
``early'' UV LFs were satisfactorily fit with Schechter functions in
line with previous UV studies. In Figure \ref{fig:fuv_lf} (upper
panel) we present an updated version of the FUV LF based on
$\sim$30,000 galaxies from \citet{s07}. All magnitudes are on AB
system and are $K$-corrected to $z=0$ and corrected for Galactic
reddening. Open points show LF of FUV absolute magnitude not corrected
for internal dust attenuation, i.e., like those of \citet{wyder} and
many other works that present {\it uncorrected} UV LFs. A standard
Schechter function is fit to the uncorrected LF (red curve). While it
visually appears as a good fit the large sample reveals that
$\chi^2_r$ is 3.7, a relatively large value. Closer look reveals that
the faint end-slope of the fit is slightly steeper than the LF points
and also that the fit falls somewhat more steeply at the bright end
($M_{\rm FUV}\lesssim -20$). \citet{wyder} were unable to identify
these discrepancies because their LF had a significantly larger error
bars and moreover, because the brightest point of their LF was at
$M_{\rm FUV} = -20$, just before the departure from the Schechter form
starts to become apparent at the bright end. Indeed,
\citet{schiminovich07}, using the same dataset as the one we use, but
analyzed independently, mention the high-end deviation from
Schechter. They tentatively ascribed it to the AGN contamination in
the UV. However, broad-line AGNs that could affect the UV continuum
were already removed in these samples, so this explanation seems
unlikely. In any case, this not-so-perfect agreement between UV LF and
Schechter function was generally neglected.

Altogether, the observed UV LF has qualitatively small departures from
the Schechter form, while given our results in \S\ 2 and 4 we would
expect a LF of SFR type (such as the UV LF) to be very poorly fit by
the Schechter function. This is because UV LF, not corrected for dust,
is not equivalent to a SFRF. Indeed, the departure from the Schechter
form becomes much more severe ($\chi^2_r=24$) when the UV LF is
constructed from {\it dust-corrected} FUV absolute magnitude, which we
show as solid dots in Figure \ref{fig:fuv_lf} (upper panel). The dust
attenuation applied in Figure 6 is obtained on galaxy-by-galaxy basis
from fitting of the full UV-optical SED \citep{s07} and is mostly
constrained by the UV slope. If, on the other hand one was to apply a
fixed dust correction, such ``corrected'' LF would simply be the
uncorrected LF shifted to brighter magnitudes, while the shape and the
steep, Schechter-like bright end would remain. Interestingly, even
applying somewhat more sophisticated statistical dust correction
brings the ``corrected'' LF only slightly closer to its true
shape. For example, many studies apply a mass or optical luminosity
dependent dust correction. From
our data on SDSS/{\it GALEX} SF galaxies we find that the medians in
mass bin yield the following relation between FUV attenuation and
stellar mass:

\begin{equation}
A_{\rm FUV} = 0.71 \log M_*-5.16.
\end{equation}
 
\noindent If we apply this statistical correction to FUV absolute
magnitudes the resulting LFs still features a relatively steep bright
end, which in LFs with smaller dynamic range could again easily be
misinterpreted as conforming to a Schechter function. The reason why
even the mass-dependent dust correction cannot reproduce the true SFRF
is because at any given mass the dispersion in dust corrections is
very large (we find 0.7 mag scatter in $A_{\rm FUV}$ at $\log
M_*=10.5$; similarly large scatter can be seen \citealt{garn10}
analysis of SDSS Balmer decrements; their Figure 4) and it is the
values that scatter above the average relation that are important in
shaping the bright end of the LF.

Guided by our previous considerations regarding the SFRF we can
attempt to fit more appropriate functions to both the uncorrected FUV
LF (where the departures from Schechter are relatively small, but not
negligible) and to the dust-corrected FUV LF (where departures are
severe). Since the {\it GALEX} sample is similar to an SFR-limited
sample as it is $M_{\rm FUV}$ limited, we try both the extended
Schechter function ( = power-law + S{\'e}rsic) and the Saunders
function (=power-law + Gaussian). We find that both functions
represent good fits to both the uncorrected and the dust-corrected FUV
LFs, but that the Saunders function is somewhat better ($\chi^2_r$ of
0.5 and 1.9 for the uncorrected and corrected LF, respectively, vs.\
$\chi^2_r$ of 0.8 and 4.5 for the extended Schechter fit). We show the
best Saunders fits as green curves in Figure \ref{fig:fuv_lf} (lower
panel). Note that the high-end tail of the uncorrected LF is now well
fit. The parameters of the best Saunders fit are:

\begin{equation}
M_{\rm FUV}^{\star}= -17.49, \quad \sigma= 0.31, \quad \gamma=-1.23,\quad \log \phi^{\star}=-2.41
\end{equation}

\noindent for uncorrected FUV LF, and:

\begin{equation}
M_{\rm FUV}^{\star}= -16.05, \quad \sigma= 0.73, \quad \gamma=-0.83, \quad \log \phi^{\star}=-2.24
\end{equation}

\noindent for the dust-corrected one.

\begin{figure}
\epsscale{1.2} \plotone{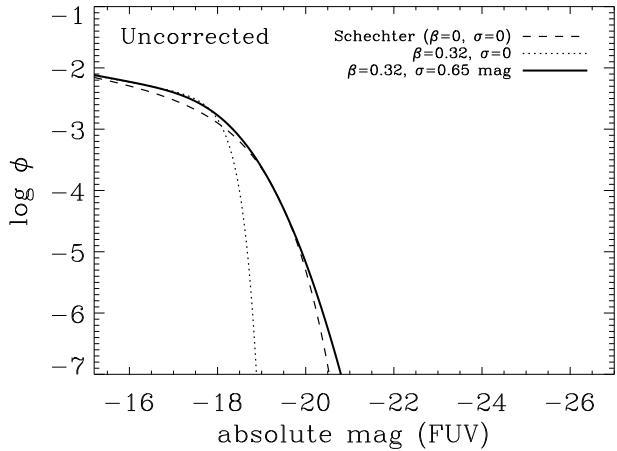}
\caption{Schematic explanation as to why the Schechter function
  appears to be an adequate description of the uncorrected FUV LF
  despite the very different character of UV and optical/near-IR
  populations. For a distribution to have a Schechter form it needs to
  have a linear dependence on mass with no scatter (dashed
  curve). Severe sub-linearity, as in FUV luminosity vs.\ mass makes
  the distribution much steeper (dotted curve), but adding the right
  amount of scatter to such non-linear relation (solid curve) modifies
  the high-end tail into a form that resembles the Schechter function
  (dashed curve). Similar principles would apply to LFs in near UV or
  the \ha\ LF.}
\label{fig:fuv_nc_coin}
\end{figure}

We now return to the apparent puzzle of why the {\it uncorrected} FUV
LF is reasonably well fit with a Schechter function (left red curve in
Figure \ref{fig:fuv_lf} (upper panel)) when our analysis has shown
that the Schechter function should be an appropriate description for
quantities that are proportional to the {\it mass} and have small
scatter with respect to it (for example, the optical or the near-IR
luminosity), which UV luminosity is not. Here we show that this near
match is a coincidence arising from two effects that approximately
cancel out -- sub-linearity of the FUV luminosity--mass relation and
the scatter in that relation.  The dust correction is on average
larger in more massive star-forming galaxies (e.g., \citealt{wang,garn10}),
therefore the slope between FUV luminosity {\it not} corrected for
dust and the stellar mass will be even lower (less linear) than
between the SFR (i.e., dust-corrected FUV luminosity) and
mass. Indeed, we determine this slope to be $\beta=0.32$ (while it was
$\beta=0.65$ for SFR, i.e. dust-corrected FUV luminosity, Eqn.\
\ref{eqn:sfr_mass}). Based on this fact alone the Schechter function
should be expected to be an even worse description for the uncorrected
FUV LF. We show this in Figure \ref{fig:fuv_nc_coin}. The dashed line
shows how the LF would look if FUV luminosity was linearly related to
mass, without any scatter. This is simply a Schechter
function. However, the actual dependency is sub-linear with
$\beta=0.32$. Such LF, but still with no scatter in L(FUV) vs.\ mass,
is shown as a dotted curve. It is much steeper at the bright end than
the Schechter function, and the high-end tail is shifted to the
left. However, we also need to take into account the {\it scatter} of
L(FUV) at a given mass. Therefore, we convolve the dotted curve with a
Gaussian with $\sigma=0.65$ mag to obtain the final LF (solid
line). Convolution has the effect of making the tail shallower again,
which {\it coincidentally} yields a distribution that resembles the
Schechter function. The differences are relatively subtle: the
slightly more shallow high-end tail and somewhat less steep faint-end
slope, and will be easily masked in LFs based on up to few thousand
galaxies. As we have seen these differences can be revealed with LFs
constructed from an order of magnitude larger samples (Figure
\ref{fig:fuv_lf} (upper panel)).

\begin{figure}
\epsscale{1.2} \plotone{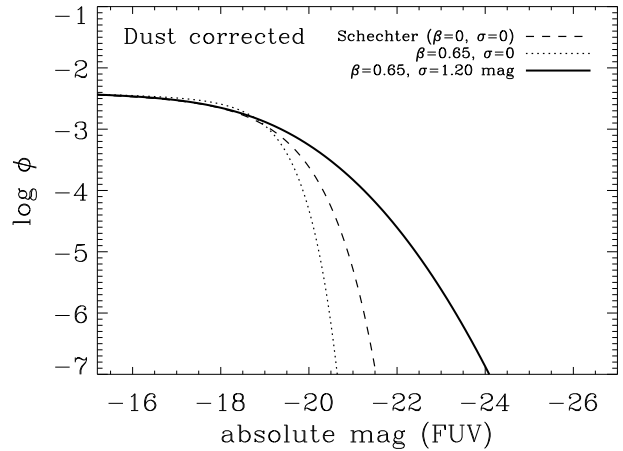}
\caption{Schematic explanation as to why the Schechter function is
  {\it not} an adequate description of the dust-corrected FUV LF
  (i.e., SFRF). The mean relation between luminosity and mass is
  closer to linear, leading to a smaller shift to the left (dotted
  curve), which results in flatter LF when scatter is applied (solid
  curve) and much larger departure with respect to Schechter function
  (dashed curve).}
\label{fig:fuv_dc_coin}
\end{figure}

This approximate cancelation of sub-linearity and scatter happens to
hold only for uncorrected luminosity. Since the dust-corrected FUV
luminosity is less sub-linear with respect to the stellar mass the
dotted line shifts less to the left. When scatter is added to it, it
moves the LF more to the right and with the shallower slope than that
of the Schechter function (Figure \ref{fig:fuv_dc_coin}), yielding
what can be considered a true SFRF and which is well described by
Saunders function.

\subsection{\ha\ and [OII] emission line LFs}

All of the considerations laid out regarding the UV LFs are also
applicable to \ha\ LFs. This is because the \ha\ luminosities also
require significant dust corrections to be representative of the true
SFRs. Ultimately, the uncorrected \ha\ LF is again only approximately
and coincidentally described by a Schechter function. Accurate \ha\
LFs do show deviations. This has been noted by \citet{gilbank} for
both \ha\ and [OII] LFs from SDSS spectra ($z\sim0.1$) and by
\citet{zhu} in their $z\sim1$ [OII] LF from DEEP2 spectra. These
studies instead fit a double power law (broken line on a log-log plot)
to the observed emission-line LFs. The dust-corrected LFs would show
even stronger departures from the Schechter form if the dust
attenuation is determined for each galaxy individually. In either case
a function such as the Saunders or the extended Schechter would
provide an optimal description.

Pa$\alpha$ is emitted in the near-IR so it requires very little dust
correction \citep{calzetti}. We anticipate that once Pa$\alpha$ LFs
are constructed, they will depart significantly from the Schechter
form and will feature a Gaussian or a S{\'e}rsic high-end tail instead.


\subsection{IR LFs}

The inadequacy of the Schechter formulation for describing IR LFs was
noticed already in the mid-80s based on {\it IRAS} 60 $\mu$m
data. Thus, \citet{lawrence} proposed a double power-law fit (i.e.,
the exponential cut off of the Schechter function was replaced by a
less steep power law), while \citet{saunders}, using an order of
magnitude larger sample (2800 vs.\ 300), noticed the curved high-end
tail of the 60 $\mu$m LF so instead recommended the combination of a
power law for the low end and the Gaussian for the high end, which we
refer to as the Saunders function (Eqn.\ \ref{eqn:saund}).\footnote{A
  number of papers published since 2005 refer to Saunders function as
  the {\it double exponential}, which we find inaccurate and
  confusing. In mathematics the double exponential refers to functions
  of the form, $f(x)=a^{b^x}$, which Saunders function is not.} The
Saunders function remains an excellent functional form for the LFs
based on the most recent reductions of the {\it IRAS} 60 $\mu$m data
containing five times as many galaxies as used in \citet{saunders},
and spanning eight orders of magnitude in space density
(\citealt{takeuchi}, see also \citealt{wangrowan}). Despite the
success of Saunders function some studies continue to use the simpler
double power law (with or without a soft transition between the two
power laws) to model the far-IR LF, especially when smaller sample
sizes ($\lesssim 10^3$) that probe smaller dynamic range of the LF are
involved (e.g., \citealt{sanders}), or when the LFs are constructed
for luminosities at $\lambda>60 \mu$m, where the measurements
typically have lower accuracy (e.g., \citealt{goto11a}). The two forms
can be rather similar, and distinguishing between them requires very
accurate LFs.

The situation appears to be similar for the mid-IR LFs. \citet{shupe}
find that the 25 $\mu$m LF constructed from IRAS data has a shallow
high-end tail and can be modeled as a double power law. Using {\it
  ISO} 15 $\mu$m measurements from ISO \citet{xu} and \citet{pozzi}
also obtained LFs with shallow high-end tails that would be
incompatible with a Schechter distribution. Given these results it
comes as a surprise that \citet{huang} found that the high end of the
$8 \mu$m LF is rather steep and is well fit by a Schechter
function. They find that this is the case even after they subtract the
stellar continuum contribution, i.e., when the resulting $8 \mu$m
emission comes primarily from PAH line emission, which is considered
to be a tracer of young populations and therefore of SF
\citep{schreiber,diaz-santos}. \citet{huang} explicitly point out that
their results are in contrast with \citet{saunders} 60 $\mu$m LF that
has excess counts above Schechter function. They also point out that
this agreement with Schechter function cannot be explained by their
removal of galaxies harboring AGN. Namely, dust-obscured AGNs could
heat the surrounding dust to relatively high temperatures, resulting
in the IR SED component that peaks in the mid-IR
\citep{fu}. \citet{huang} offer no explanation for the puzzling
inconsistency between their 8 $\mu$m PAH LF and LFs at longer IR
wavelengths. We suggest that the reason behind this is because the PAH
emission is perhaps not a good tracer of the {\it current} SF. Namely,
there are indications that the PAHs exist outside of HII regions where
they are heated by the general interstellar radiation field produced
by {\it older} stars \citep{calzetti11}. By tracing intermediate age
($\sim 1$ Gyr) or older stellar populations PAH luminosity immediately
becomes more closely related to the {\it stellar mass} and less so to the
current SF, and thus PAH luminosity function can be expected to more
closely follow the MF and therefore the Schechter distribution. In
other words, PAH LF is perhaps not a true SFRF.

The curious fact that some LFs of SFR type (primarily UV and \ha\ LFs)
are Schechter-like, which we now explain to be unrelated to the
Schechter form of LFs of mass type, and the consequent uncertainties
as to the true shape of SFRF left a vacuum that some recent studies
try to fill by proposing that the IR LF function is intrinsically also
of Schechter form, but that some effect not related to SF makes the
tail assume shallower non-Schechter slope. These studies find a
culprit among the dust-obscured AGN, which, as mentioned, can affect
the IR SED, especially in the mid IR. If correct, such explanation
would disagree with our conclusion (and the view of many previous IR
studies, e.g., \citealt{takeuchi,takeuchi10a,buat07,buat09}) that the
IR LF, to the extent that it measures current SF, and correspondingly
the SFRFs, are intrinsically not of Schechter form.

We first discuss the evidence for AGN contamination in the mid IR
(where it is expected to be stronger) and then the far IR (which also
dominates the total IR luminosity). To test the AGN contamination
hypothesis one needs to remove the AGN contribution to the IR
luminosities and construct SF-only IR LF. \citet{fu} used {\it
  Spitzer} IRS spectra of $z\sim0.7$ galaxies to construct LFs at 8
and 15 $\mu$m rest-frame. Mid-IR spectroscopy allowed them to
decompose SF and AGN components on a galaxy by galaxy basis (at least
for the galaxies in the bright tail). AGNs were found to dominate at
the highest 8 and 15 $\mu$m luminosities. After removing their
contribution they find that the resulting LFs are well fit by
Schechter functions. We note that \citet{fu} probe LFs with good
precision over only 1 dex of luminosities, so the shapes of the
resulting LFs are not well determined.  Following in the footsteps of
\citet{fu}, \citet{wu} use mid-IR spectra to decompose star-forming
and AGN contributions at rest frame 15 and 24 $\mic$ for a more local
sample ($z<0.3$). They confirm that after correcting for AGN
contribution the 15 and 24 $\mic$ LFs become formally consistent with
the Schechter functions, however, we notice that their data do not
probe the high end sufficiently well to rule out non-Schechter
distribution.\footnote{For example, as shown in their Fig.\ 7 even
  before the AGN correction their 15 $\mic$ LF can be fitted by either
  Schechter or Saunders functions. Unfortunately, one LF point that
  could help understand what happens at the high end was omitted from
  their SF LF in Fig.\ 8a.} Interestingly, \citet{rujopakarn} who
instead of decomposing the AGN contributions completely exclude them
find much smaller difference between the total and SF-only 24 $\mic$
LFs. Consequently, they fit both with double power laws and do not
consider Schechter function at all. To conclude, there is tentative
evidence that after the AGN correction the mid-IR LF is
Schechter-like. This possible agreement of the mid-IR LFs with the
Schechter distribution could be explained if the mid-IR continuum does
not truly trace the current SFR. Indeed, \citet{kelson} present a
model in which most of the mid-IR emission is produced by carbon AGB
stars with ages between 0.2 and 2 Gyr, i.e., while \citet{s09} find
that the $\sim 15 \mic$ luminosity is more tightly correlated with the
optical $B$-band luminosity, where intermediate age stars dominate,
than it is with the dust-corrected FUV luminosity of young
stars. Significant contribution of low mass stars in the heating of
mid-IR dust would make it be more strongly correlated with the
cumulative SF and therefore the stellar mass, than to the current
SF. Consequently, the mid-IR LFs may be of mass function type and thus
be better described by Schechter functions.

AGN contribution can be accessed relatively directly in the mid-IR
where the AGN SED peaks, but it is more uncertain in the far and the
total IR. \citet{wu}, extrapolating from their 24 $\mic$ results, find
that the AGN contribution to the total IR luminosity is only 10 to
20\% (out to$\log L_{\rm IR}=11.7$), yet some studies attempt to
correct for AGN contribution in the total IR luminosity. This can be
done in two ways. One is to keep all galaxies but remove fraction of
IR luminosity believed to come from an AGN, and the other, more crude
method is to remove galaxies showing signs of AGN from the LF
altogether.  The IR LF constructed using former method is equivalent
to the SFRF, while the latter will be a lower limit to the true
SFRF. The recent example of the IR LF that completely removes galaxies
with AGNs is the one presented in \citet{goto11a}, derived from {\it
  AKARI} data. \citet{goto11a} show full IR LF alongside one that
removes AGNs based on the {\it number} fractions of optically
identified AGNs in each IR luminosity bin from
\citet{yuan-kewley}. \citet{yuan-kewley} AGN number fraction is a
steeply rising function of IR luminosity. Thus the \citet{goto11a} LF
of non-AGN galaxies has a steeper bright end than the full LF, but
still not so steep to make the authors consider fitting the Schechter
function instead of the double power law. We confirm this by
performing the fits ourselves on \citet{goto11a} LFs. Schechter
functions yield very poor fit for either the total or just the non-AGN
LF ($\chi^2_r= 10$ and 11 respectively). On the other hand, the
Saunders functions produce excellent fits with $\chi^2_r= 0.6$ and 0.7
for the total and non-AGN LFs respectively. Since the non-AGN LF is
only the lower limit to the real SF IR LF, the latter also cannot be a
Schechter function. The conclusion that the AGNs are not responsible
for the non-Schechter form of IR LF is also in line with the exquisite
LF at 60 $\mu$m from {\it IRAS} data. \citet{takeuchi} show that the
60 $\mu$m LF does ha an excess at $L_{60}>10^{11.5} L_{\odot}$, but
this excess is only $\approx 0.2$ dex in $L_{60}$ and lies above the
{\it Saunders} fit and not just above a putative steep Schechter
high-end slope.

Finally, another strong observational confirmation that the true SFRF
is not of Schechter form comes from the LFs of radio galaxies, where
the separation between AGNs and the star forming component can be
achieved more easily than in the IR. Thus, for example, \citet{mauch}
show that the 1.4 GHz LF of star-forming radio galaxies is very well
fit by a Saunders function, but not Schechter.

To summarize, while there seems to be some evidence that the {\it mid}
IR LFs after correcting for AGN contribution may be consistent with a
Schechter function, we propose that this could be the consequence of
the mid-IR luminosity intrinsically tracing SF over longer timescales,
and therefore being more related to the stellar mass than to the {\it
  current} SF. On the other hand, it appears that for either the
far-IR or the total IR even with very liberal AGN corrections do not
produce Schechter-like LFs. On the contrary, far and total IR LFs
remain well described by functions that replace the exponential cutoff
of Schechter function with less steep functional forms, especially the
Gaussian (as featured in Saunders function), in line with the
expectations from our simulations.

\subsection{Distribution of specific SFRs}

Specific SFR (sSFR) normalizes SFR by stellar mass of a galaxy, thus
allowing us to asses its SF history and characterize it as bursty,
normal or quiescent. The distribution of specific SFRs (sSFR function)
can offer complementary insights to those offered by SFRF alone. We
have already shown that at any given mass the observed SFR
distribution is well described by two Gaussians: one for the SF
sequence and another for a broad, passive sequence. This double
Gaussianity is preserved when the bimodal SFR--mass relation is
projected to produce a mass-limited SFR distribution. The same is true
for the distribution of {\it specific} SFRs. It too can be described
as the composite of two Gaussians in log (SFR/$M_*$).

Recently, \citet{sargent} have proposed that at $z\sim2$ what we call
the SF sequence and model as a single Gaussian is in itself composed
of {\it two} Gaussians -- one in which the majority of normal SF
galaxies lie, and another one, forming a bump on the side of the main
one, that contains strongly starbursting galaxies. Furthermore,
\citet{sargent} present a framework in which this two-mode sSFR
distribution extends to lower redshifts, and they show that it can
explain the non-Schechter character of the local IR LF.\footnote{To be
  accurate, \citet{sargent} do not construct a sSFR distribution but a
  sSFR distribution relative to the peak of the SF sequence. The two
  are very similar.} However, our analysis of the sSFR distribution of
40,000 $z\sim0.1$ galaxies, performed in exactly the same way, shows
absolutely no indication that the SF sequence has a second,
starbursting mode. Most likely this mode has became negligible since
$z\sim2$. More importantly in the context of this study is that no
second mode in the SF sequence is needed to produce an IR LF (i.e.,
SFRF) with non-Schechter form. As shown in \S\ 2.3., the non-Schechter
distribution is primarily the result of a scatter in the SFR--mass
relation within the unimodal SF sequence alone.

\subsection{Bivariate (s)SFR--$M_*$ distributions}

The approach in this work was to arrive at SFRFs by first modeling the
SFR vs.\ $M_*$ distribution. SFRF collapses the information from the
bivariate SFR--$M_*$ distribution. Therefore, our recommendation is
that all studies that report on SFRFs or LFs of SFR type should also
construct a SFR (or sSFR) vs.\ $M_*$ diagram (even a simple scatter
plot) to aid in the interpretation of the SFRF. From such diagram one
should try to determine the slope and the scatter of the SF sequence
and get some sense of the fraction of galaxies on the passive
sequence, as well as determine the actual mass and SFR limits. For
most purposes even crude mass estimates would suffice. Techniques
developed for the construction of bivariate LFs
\citep{takeuchi10b,takeuchi12,johnston} can also be applied for the
construction of formal bivariate SFR--$M_*$ and sSFR--$M_*$
distributions.

\section{Conclusions}

The main conclusions of this study can be summarized as the following:

\begin{enumerate}

\item Distributions of the SFR and the stellar mass are fundamentally
  different. Consequently, the LFs related to mass (optical and
  near-IR) will differ from dust-corrected LFs related to SF (UV,
  emission line, IR).

\item SFR distributions (SFR functions) are very poorly described by a
  Schechter functional form, which is adequate for mass
  functions. Instead, SFR functions (of SFR-limited samples) are very
  well described by either an extended Schechter function (which
  replaces the exponential cutoff with a S{\'e}rsic function) or by a
  Saunders function (where the high end is described by a Gaussian in
  log SFR); see Figures \ref{fig:bim} (lower right panel) and
  \ref{fig:fuv_lf} (lower panel). Both feature four parameters. In
  several empirical cases that we tested (our far-UV LF and IR LFs of
  \citealt{goto11a} and \citealt{takeuchi}) the Saunders function
  produced somewhat better fits. Saunders function has an additional
  advantage that its parameters are less covariant between each other
  and are therefore more robust to LF measurement errors.

\item SFR functions of mass-limited samples feature a drop at the low
  end even for volume-complete samples. This drop should not be
  confused with incompleteness. The shape of SFR functions at the low
  end is critically sensitive to the presence of mass
  limits. Mass-limited SFR functions are well described with double
  Gaussians in log SFR (Figures \ref{fig:bim} (lower left panel) and
  \ref{fig:obs}).

\item As previous studies have shown, the observed UV (and \ha) LFs
  with no dust correction (or with average statistical dust
  corrections) can approximately be described by a Schechter function
  (Figure \ref{fig:fuv_lf} (upper panel)). Schechter form in them is
  not fundamental (like it is in optical LFs) but is a consequence of
  two effects that by chance approximately cancel each other out
  (Figure \ref{fig:fuv_nc_coin}). Precise LFs (using samples with
  $>10^4$ galaxies) reveal that even the uncorrected UV LF deviates
  from Schechter function. UV and \ha\ LFs need to be dust corrected
  on galaxy-by-galaxy basis and need to have at least a moderate
  dynamic range for departures from Schechter form to be
  evident. These requirements are not always fulfilled in
  high-redshift studies, leading to apparent agreements of UV and \ha\
  LFs with the Schechter form. When properly dust corrected, UV LFs
  follow our mock SFRFs and are successfully described by Saunders
  functions.

\item LFs in the far and total IR, as well as the radio LF for star
  forming galaxies, behave like our mock SFRFs and are poorly fit
  with a Schechter function even when all AGNs are removed. Instead,
  they are very well fit by Saunders functions.

\item LFs in the {\it mid}-IR may intrinsically be Schechter-like
  (after correcting for AGN contribution), which could be the
  consequence of the mid-IR tracing less massive (older) stellar
  populations; i.e., mid-IR LFs are possibly not real SFR functions
  but are more closely related to stellar mass functions. Therefore,
  Schechter functions would represent an adequate description.

\item Whenever possible a bivariate SFR--$M_*$ distribution should be
  considered alongside its projections.

\end{enumerate}

The recognition that optical LFs and MFs can be characterized using a
functional form proposed by \citet{schechter} has provided extremely
valuable guidance in the study of galaxy populations across a range of
redshifts. As our estimates of SFRs have improved in recent years,
having a similar tool to apply to SFR distributions will hopefully add
to our understanding of galaxies and their evolution.

\acknowledgments

S.S.: ``I dedicate this paper to the memory of my mother Mirjana
Makra-Salim (1947-2012). I am forever grateful for her selfless love.''

\end{document}